\documentclass{aa}  
\usepackage{graphicx}
\usepackage{amsmath}	% Advanced maths commands
\usepackage{amssymb}	% Extra maths symbols
\usepackage{relsize}
\usepackage{txfonts}
\usepackage[colorlinks=true,linkcolor=blue,allcolors=blue]{hyperref}%
\newcommand{\Msun}{M$_\odot$}

\newcommand{\Lsun}{L$_\odot$}

\newcommand{\MLsun}{M$_\odot$/L$_\odot$}
\newcommand{\magarcsq}{$\mathrm{mag\,arcsec^{-2}}$}

\newcommand{\Mstar}{\ensuremath{M_*}}

\newcommand{\rcut}{\ensuremath{r_\mathrm{cut}}}

\newcommand{\Mshalo}{\ensuremath{M_{*,\mathrm{halo(r< 2R_e)}}}}

\begin{document} 

   \title{The Fornax3D project: discovery of ancient massive merger events in the Fornax cluster galaxies NGC\,1380 and NGC\,1427}

   \subtitle{}

   \author{
   % main
   Ling Zhu$^1$\thanks{E-mail: lzhu@shao.ac.cn}, Glenn van de Ven$^{2}$, Ryan Leaman$^{2,3}$, Annalisa Pillepich$^3$, 
   % significant contributors, alphabetical
   Lodovico Coccato$^{4}$, Yuchen Ding$^{1}$, Jes\'us Falc\'on-Barroso$^{5,6}$, Enrichetta Iodice$^{4}$, Ignacio Martin Navarro$^{5,6}$, Francesca Pinna$^{3}$, 
   % others, alphabetical
   Enrico Maria Corsini$^{7,8}$,  Dimitri A. Gadotti$^{4}$, Katja Fahrion$^{4,9}$, Mariya Lyubenova$^{4}$, Shude Mao$^{10}$, Richard McDermid$^{11}$, Adriano Poci$^{11}$, Marc Sarzi$^{12,13}$, Tim de Zeeuw$^{14,15}$}

   \institute{
Shanghai Astronomical Observatory, Chinese Academy of Sciences, 80 Nandan Road, Shanghai 200030, China\\
\email{lzhu@shao.ac.cn}
\and
Department of Astrophysics, University of Vienna, T\"urkenschanzstra{\ss}e 17, 1180 Vienna, Austria
\and
Max Planck Institute for Astronomy, K\"onigstuhl 17, 69117 Heidelberg, Germany % MPIA
\and
European Southern Observatory, Karl-Schwarzschild-Stra\ss{}e 2, 85748 Garching bei M\"unchen, Germany %ESO Lodo, Enrica, Dimitr, Katja, Mariya
\and
Instituto de Astrof\'isica de Canarias, Calle Via L\'{a}ctea s/n, 38200 La Laguna, Tenerife, Spain% Jesus
\and
Depto. Astrof\'isica, Universidad de La Laguna, Calle Astrof\'isico Francisco S\'{a}nchez s/n, 38206 La Laguna, Tenerife, Spain% Jesus
\and
Dipartimento di Fisica e Astronomia 'G. Galilei', Universit\`a di Padova,  vicolo dell'Osservatorio 3, I-35122 Padova, Italy %Enrico
\and
INAF--Osservatorio Astronomico di Padova, vicolo dell'Osservatorio 5, I-35122 Padova, Italy % Enrico
\and
European Space Agency (ESA), European Space Research and Technology Centre (ESTEC), Keplerlaan 1, 2201 AZ Noordwijk,
  The Netherlands % Katja
 \and
Department of Astronomy and Tsinghua Center for Astrophysics, Tsinghua University, Beijing 100084, China % Dandan, Shude
\and
Department of Physics and Astronomy, Macquarie University, North Ryde, NSW 2109, Australia %Richard
\and
Armagh Observatory and Planetarium, College Hill, Armagh, BT61 9DG, Northern Ireland, UK  % Sarzi
\and
Centre for Astrophysics Research, University of Hertfordshire, College Lane, Hatfield AL10 9AB, UK % Sarzi
\and
Sterrewacht Leiden, Leiden University, Postbus 9513, 2300 RA Leiden, The Netherlands % Tim
\and
Max-Planck-Institut f\"ur extraterrestrische Physik, Gie\ss{}enbachstra\ss{}e 1, 85748 Garching bei M\"unchen, Germany. % Tim
             }

   \date{Received; accepted}
   
   \titlerunning{Discovery of ancient massive merger events}
\authorrunning{Zhu et al.}  
 
  \abstract
   {
We report the discovery of ancient massive merger events in the early-type galaxies NGC\,1380 and NGC\,1427, members of the Fornax galaxy cluster.
Both galaxies have been observed by the MUSE IFU instrument on the VLT, as part of the Fornax3D project. 
By fitting recently-developed population-orbital superposition models to the observed surface brightness as well as stellar kinematic, age, and metallicity maps, we obtain the stellar orbits, age and metallicity distributions of each galaxy. 
We then decompose each galaxy into multiple orbital-based components, including a dynamically hot inner stellar halo component which is identified as the relic of past massive mergers. 
By comparing to analogues from cosmological galaxy simulations, chiefly TNG50, we find that the formation of such a hot inner stellar halo requires the merger with a now-destroyed massive satellite galaxy of $3.7_{-1.5}^{+2.7} \times 10^{10}$\,\Msun\ (about $1/5$ of its current stellar mass) in the case of NGC\,1380 and of $1.5_{-0.7}^{+1.6} \times10^{10}$\,\Msun\ (about $1/4$ of its current stellar mass) in the case of NGC\,1427. 
Moreover, we infer that the last massive merger in NGC\,1380 happened $\sim10$ Gyr ago based on the stellar age distribution of the re-grown dynamically cold disk, whereas the merger in NGC\,1427 ended $t\lesssim 8$ Gyr ago based on the stellar populations in its hot inner stellar halo. 
The major merger event in NGC\,1380 is the first one with both merger mass and merger time quantitatively inferred in a galaxy beyond the Local Volume. Moreover, it is the oldest and most massive merger uncovered in nearby galaxies so far.
}

\keywords{galaxies: structure -- galaxies: dynamics -- galaxies:observations  -- galaxies: stellar kinematics}

   \maketitle
%
%-------------------------------------------------------------------

\section{Introduction}

Driven by gravity, galaxies are expected to continuously grow through the merging of smaller systems. Deriving their past merger history is challenging, as the accreted stars disperse quickly. Uncovering the properties of past merger events was so far only possibly in the nearest few galaxies for which individual stars can be resolved in observations.

In our own galaxy, the Milky Way, motion and chemistry of individual stars are detected. Numerous streams and over-densities \citep[e.g.,][]{Majewski2003, Helmi2020, Ibata2021}
identified in the outer halo indicate the many minor mergers still on-going in the Milky Way. Debris from massive ancient mergers is spatially well-mixed, but fortunately still distinguishable in kinematics.
Recently, stars on highly-radial orbits in the vicinity of the Sun have been associated to the Galaxy's inner stellar halo -- e.g. the so-called Gaia Enceladus or Gaia Sausage, which is thought to have originated from a single massive merger of $\sim3-6 \times 10^8$\,\Msun\ about $10$\,Gyr ago \citep{Helmi2018, Belokurov2018}. 

In our neighboring galaxy, the Andromeda giant spiral galaxy (M31), an inner stellar halo with a giant stream has been identified \citep{Ibata2014}, which could have been also induced by a single massive merger, called M32p, with $\sim2\times10^{10} $\,\Msun\ about $\sim2$\,Gyr ago \citep{DSouza2018, Hammer2018}. The satellite mass is inferred from the metallicity of M32, which is supposed to be the surviving core of the progenitor satellite galaxy. The merger time was inferred from stellar age distribution of stars in the giant streams and inner stellar halo.

In the nearest giant elliptical galaxy, NGC 5128 at distance of 3.8\,Mpc, individual stars in the outer halo are resolved by deep HST photometry \citep{Rejkuba2011} and analysis similar to that of M31 indicates a past massive merger of also $\sim2\times 10^{10}$\,\Msun\, \citep{DSouza2018b}.

In the above cases, the quantification of the satellite mass relies on the mass-metallicity relation, while the constraint on merger time relies on star formation histories obtained from individual stars. Similar studies are thus limited to the nearest few galaxies until 30-m telescopes become available and allow us to get resolved stellar populations for more distant galaxies.
The past merger histories for a few galaxies at larger distances are partially uncovered using different techniques. Individual GCs in the halos are detected in nearby galaxies and are used as another tracer of the galaxy assembly history \citep{Forbes2016, Beasley2018,Kruijssen2019}. 
Deep images broadly characterize the light distribution of stellar halos in nearby galaxies and found varies substructures, like streams, in the halo \citep{Merritt2016, Harmsen2017}.

Over the last two decades, integral-field-unit (IFU) instruments have spectropscopically mapped thousands of galaxies across a wide range of masses and Hubble types \citep{dezeeuw2002, Cappellari2011, Sanchez2012, Bryant2015, Bundy2015}.
In principle, the information of stellar motions and chemical distributions of an external galaxy is included in these IFU data, though all mixed along the line-of-sight. 
Recent studies have attempted to constrain the global ex-situ fractions \citep{Davison2021b, Davison2021, Boecker2020} or the mass of satellite mergers \citep{Pinna2019, Pinna2019b, Martig2021} based on age and metallicity distributions of stars in galaxies' inner regions obtained from IFU data.
These methods can identify minor mergers as their accreted stars have lower metallicity than in-situ stars, but they become insenstive to major mergers with overlapping metallicities. Also, the merger times remain largerly unconstrained with such approaches.

From a numerical perspective, recent and current cosmological hydrodynamical simulation projects, such as IllustrisTNG\footnote{\url{www.tng-project.org}} and EAGLE\footnote{\url{http://eagle.strw.leidenuniv.nl/}}, have been able to reproduce large numbers of galaxies across the morphological spectrum with well-resolved structures \citep{Pillepich2019, Pulsoni2020, Rodriguez-Gomez2019, Du2019, Correa2017}. 
Structures like Gaia-Enceladus in the Galactic inner stellar halo also arise in Milky Way-like simulated galaxies \citep[e.g.][]{Grand2020}. 
In \cite{Zhu2021}, a hot inner stellar halo, consisting of orbits which are dominated by random instead of ordered motion, is defined universally for all type of galaxies across a wide mass range. The mass of such a hot inner stellar halo is found to be strongly correlated with the stellar mass of the most massive merger as well as with the total ex-situ stellar mass. The correlations are consistent across the TNG50 \citep{Pillepich2019, Nelson2019}, TNG100 \citep{Pillepich2018b, Nelson2018, Springel2018,Marinacci2018,Naiman2018} and EAGLE \citep{Schaye2015, Crain2015} simulations for galaxies with stellar mass $\gtrsim4\times10^{10}$\,\Msun, regardless of the numerical resolution and galaxy formation model. The hot inner stellar halo is thus a promising way to quantify the past merger mass for all types of massive galaxies.

In parallel, we have been able to accurately infer the internal stellar orbit distribution of external galaxies by applying Schwarzschild's orbit-superposition dynamical method to the stellar kinematic maps, extracted from IFU data \citep{vdB2008,Zhu2018a}. The method has been applied to hundreds of galaxies across the Hubble sequence \citep{Zhu2018c,Zhu2018b,Jin2020,Santucci2022}.
Stellar orbits in the model have been further colored with age and metallicities in a recently-developed population-orbit superposition method \citep{Poci2019,Zhu2020}. By fitting the stellar kinematic, age and metallicity maps simultaneously, we have been able to obtain the internal stellar orbit distribution as well as the age and metallicity distributions \citep{Poci2021}.
With this novel method, chemo-dynamically identification of galactic structures as employed in the Milky Way, has become possible for external galaxies. 

In this paper, we show that the mass of the hot inner stellar halo is a quantity that can be robustly obtained for external galaxies by applying our population-orbit superposition method to VLT/MUSE IFU data of two Fornax cluster galaxies, NGC\,1380 and NGC\,1427, observed as part of the Fornax3D project \citep{Sarzi2018}. We subsequently contrast these robust mass measurements with the correlations obtained from simulations as presented in \citet{Zhu2021} to infer the masses of the mergers that induced the hot inner stellar halos in both galaxies. Finally, we use the modelled age distributions of the different dynamical structures to also constrain the times of the last major mergers in NGC\,1380 and NGC\,1427.

The paper is organized as follows. In Section~\ref{S:data}, we describe the observational data of NGC\,1380 and NGC\,1427. In Section~\ref{S:method}, we present the population-orbit superposition models of the two galaxies and measure the mass of their hot inner stellar halos. 
In Section~\ref{S:results}, we present the main results of the paper and after discussing effects due to the cluster environment in Section~\ref{S:discussion}, we summarize in Section~\ref{S:summary}. Tests of our method and evaluation of model uncertainties are outlined in the Appendices~\ref{A:method}~and~\ref{A:uncertainties}.

%---------------------------------------------------------------------
\section{Deep imaging and spectroscopic data}
\label{S:data}
% ---------------------------------------------------------------------
%---------------------------------------------------------------------
\subsection{NGC\,1380 and NGC\,1427 in the Fornax cluster}
% ---------------------------------------------------------------------
Astronomers have observed galaxies in the Fornax cluster in great detail in recent years. 
The Fornax Deep Survey (FDS) has observed all galaxies in the 9 square degrees around the core of the Fornax cluster with the VLT Survey Telescope down to 27 \magarcsq\ in $r$-band \citep{Iodice2019a}. 
Then, the Fornax3D project \citep{Sarzi2018}, using the MUSE IFU instrument on the VLT, has observed all 23 early-type galaxies (ETGs) and 10 late-type galaxies (LTGs) within the virial radius of the cluster, down to 25 \magarcsq\ in the $B$-band. 
NGC\,1380 (FCC 167) and NGC\,1427 (FCC276) are two galaxies in the Fornax cluster observed as part of the FDS \citep{Iodice2019a} and Fornax3D survey \citep{Sarzi2018}. 
 
NGC\,1380 is a lenticular galaxy located on the north-western side of the cluster at $\sim 40$\,arcmin from the brightest central galaxy (BCG) NGC\,1399, and at a distance of 21.2\,Mpc from us \citep{Blakeslee2009}. NGC\,1380 has a total $r$-band magnitude of $m_r = 9.27$ and a half-light radius of $R_e = 56.2\arcsec = 5.8$\,kpc. Deep imaging reveals an outer and fainter isophotoes at $R>3$\,arcmin, which appear less flattened and twisted with respect to the inner and more disky regions \citep{Iodice2019a}. 
Through the Fornax3D project, NGC\,1380 was observed with three MUSE pointings out to a galactocentric radius of $150\arcsec$ along the major axis and up to $40\arcsec$ along the minor axis (Figure~\ref{fig:bestfit}). It is one of the most-studied objects within the cluster. Some initial dynamical modelling results have been included in \citet{Sarzi2018}. 

NGC\,1427 is the brightest elliptical galaxy located on the eastern side of the cluster at $\sim 50$\,arcmin from the BCG, and at a distance of 19.6\,Mpc from us \citep{Blakeslee2009}. NGC\,1427 has $m_r =10.15 $ mag and $R_e =44.7\arcsec=4.2$\,kpc. This galaxy becomes bluer with increasing radius and in the outskirts, and its outer and fainter isophotes appear asymmetric at $R\gtrsim 3$\,arcmin \citep{Iodice2019a}. NGC\,1427 was observed with two MUSE pointings out to a radius of $100\arcsec$ along the major axis and up to $40\arcsec$ along the minor axis (Figure~\ref{fig:bestfit_1427}).

Both NGC\,1380 and NGC\,1477 are located beyond the visible extended stellar halo of the BCG NGC\,1399, which is mapped out to 33 arcmin \citep{Iodice2016}.

%---------------------------------------------------------------------
\subsection{Stellar kinematic, age and metallicity maps}
% ---------------------------------------------------------------------
The IFU spectra were spatially binned using a Voronoi tessellation\citep{Cappellari2003} to reach a minimal signal-to-noise $S/N=100$ per bin, resulting in a total of 4300 bins for NGC\,1380 and 3072 bins for NGC\,1427.
Following earlier analysis of NGC\,1380 \citep{Sarzi2018}, the stellar kinematics were extracted by applying pPXF full-spectral fitting \citep{Cappellari2004} to the wavelength range 4750-5500 \AA.
This yields high-quality maps of the stellar mean velocity $V$, velocity dispersion $\sigma$, and higher order velocity moments parameterised through the Gauss-Hermite (GH) coefficients $h_3$ and $h_4$.

To obtain high-quality maps of stellar age and metallicity, the spectra were spatially re-binned to reach $S/N=200$ for NGC\,1380, resulting in a total of 974 bins. For NGC\,1427, we kept the 3072 bins with target $S/N=100$ from above.  
Each binned spectrum was fitted with regularized pPXF \citep{Cappellari2017} using the MILES single stellar population (SSP) models \citep{miles2011} based on BaSTI isochrones \citep{Vazdekis2015}, and we assume a constant Milky Way-like Kroupa \citep{kroupa2002} stellar initial mass function. 
The models cover a wavelength range of 3540-7410 \AA, and are sampled at a spectral resolution of 2.51 \AA\, \citep{Falcon2011}. They include 8 values of total metallicity [M/H]=[-0.96, +0.40] dex (with a resolution between 0.14 and 0.48 dex), 33 values of age between 1.0 and 14.0 Gyr (with a resolution between 0.25 and 0.5 Gyr), and two values of [Mg/Fe]=0.0 (solar abundance) and 0.4 dex (supersolar). The templates with metallicity below -0.96 and age below 1.0 Gyr were excluded. We fitted the stellar kinematics and populations in the same run, using a multiplicative polynomial of 8th order and regularization parameter of 0.5. We generally followed the same approach as earlier work \citep{Pinna2019} but re-normalized the templates to get light-weighted stellar age $t$ and metallicity $Z/Z_\odot$ in each Voronoi bin. 

%---------------------------------------------------------------------
\section{Population-orbital superposition method}
\label{S:method}
%---------------------------------------------------------------------
We fit population-orbit superposition models to the luminosity distribution, stellar kinematic, age, and metallicity maps for each galaxy.
The major steps of constructing the model include:(1) construction of the  gravitational potential, (2) calculation of an orbit library in the gravitational potential, (3) fitting to the luminosity density and kinematic maps by weighting the orbits, (4) fitting to the age and metallicity maps by coloring the orbits. Best-fitting models are obtained by exploring the free parameters in the gravitational potential and minimizing the $\chi^2$ difference between models and data.
In the end, we obtain 3D models superposed by stellar orbits tagged with age and metallicity, which match all data well. We then decompose the galaxies based on the modelled stellar orbit distribution. 

While our method was introduced and extensively verified in \citet{Zhu2020}, we present the main steps below.  

%---------------------------------------------------------------------
\subsection{Gravitational potential}
%---------------------------------------------------------------------
The gravitational potential is assumed to be generated by the stellar mass plus dark matter halo. Also a central black hole of $10^{8}$\,\Msun\, is added although it remains unresolved by the kinematic data and does not affect our results. 
To obtain the stellar mass distribution, we first fit the $r$-band image from FDS by a Multiple Gaussian Expansion (MGE) model \citep{Cappellari2002}. 
Then by adopting a set of viewing angles ($\vartheta, \psi, \phi$), we de-project toward a triaxial MGE model which represents the intrinsic stellar luminosity density. 
Finally, by multiplying with a constant stellar mass-to-light ratio $M/L_r$, we arrive at the intrinsic stellar mass distribution. The effects of the constant stellar mass-to-light ratio on our results are discussed in Appendix~\ref{A:uncertainties}.

The three viewing angles relate directly to three axis ratios $(p,q,u)$ describing the intrinsic shape \citep{vdB2008}.
We leave the intrinsic intermediate-to-major and minor-to-major axis ratios $(p,q)$ free to vary, but limit $u$ in the range of $0.98-0.9999$. Moderate triaxiality is allowed in the model.
The dark halo is throughout assumed to be spherical with a NFW \citep{nfw1997} radial profile with two free parameters: the halo (virial) mass $M_{200}$, defined as the total mass enclosed within the virial radius $r_{200}$ within which the average density is 200 times the critical density, and the concentration $c$, defined as the ratio between dark matter virial radius and the dark matter scale radius. We let $M_{200}$ free and for each dark matter halo we fix $c$ following the relation from \citet{Dutton2014}.
In total we thus have five free so-called hyper-parameters: $M/L_r$, $p$, $q$, $u$, and $M_{200}$.

%---------------------------------------------------------------------
\subsection{Fitting the data}
%---------------------------------------------------------------------

%---------------------------------------------------------------------
\subsubsection{Fitting to luminosity density and stellar kinematic maps}
%---------------------------------------------------------------------

%%%FIG
\begin{figure*}
\centering\includegraphics[width=12cm]{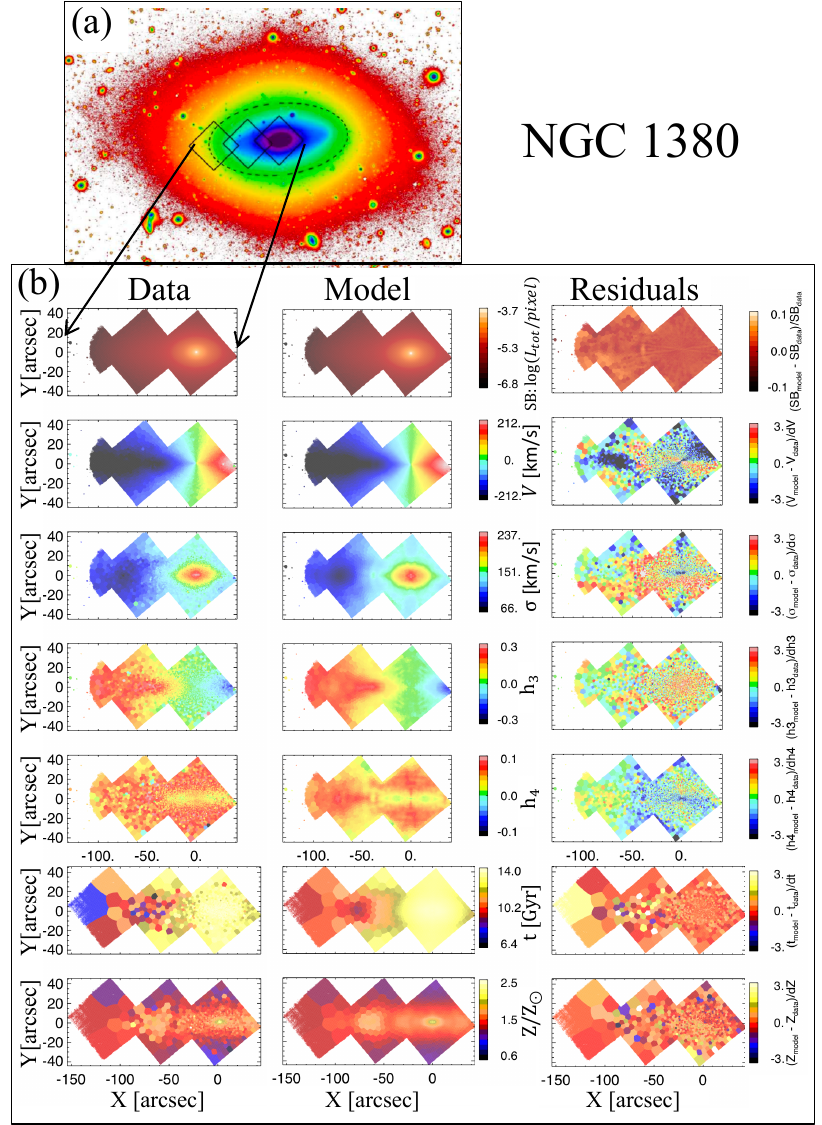}
\caption{
\textbf{Population-orbit model of NGC\,1380.}
Panel (a): A deep $r$-band image of NGC\,1380 from the Fornax Deep Survey with the VLT Survey Telescop \citep{Iodice2019a}. The three squares indicate the three pointings, each with $1'\times 1'$ square field-of-view, of spectroscopic observations with the VLT MUSE instrument. The black dashed ellipse corresponds to the isophote at $\mu_B = 25$ mag arcsec$^{-2}$.
Panel (b): The first column shows the data from the VLT/MUSE observations, the second column the best-fitting population-orbit model, and the third column the residuals of data minus model divided by the observational errors. From top to bottom: surface brightness, mean velocity, velocity dispersion, Gauss-Hermite coefficients $h_3$ and $h_4$, as well as light-weighted age and metallicity maps of the stars in NGC\,1380. Note that kinematic maps were derived from spectra binned with a different scheme than the age and metallicity maps. 
}
\label{fig:bestfit}
\end{figure*}
%%%FIG
%\note{hi

\begin{figure*}
\centering\includegraphics[width=10cm]{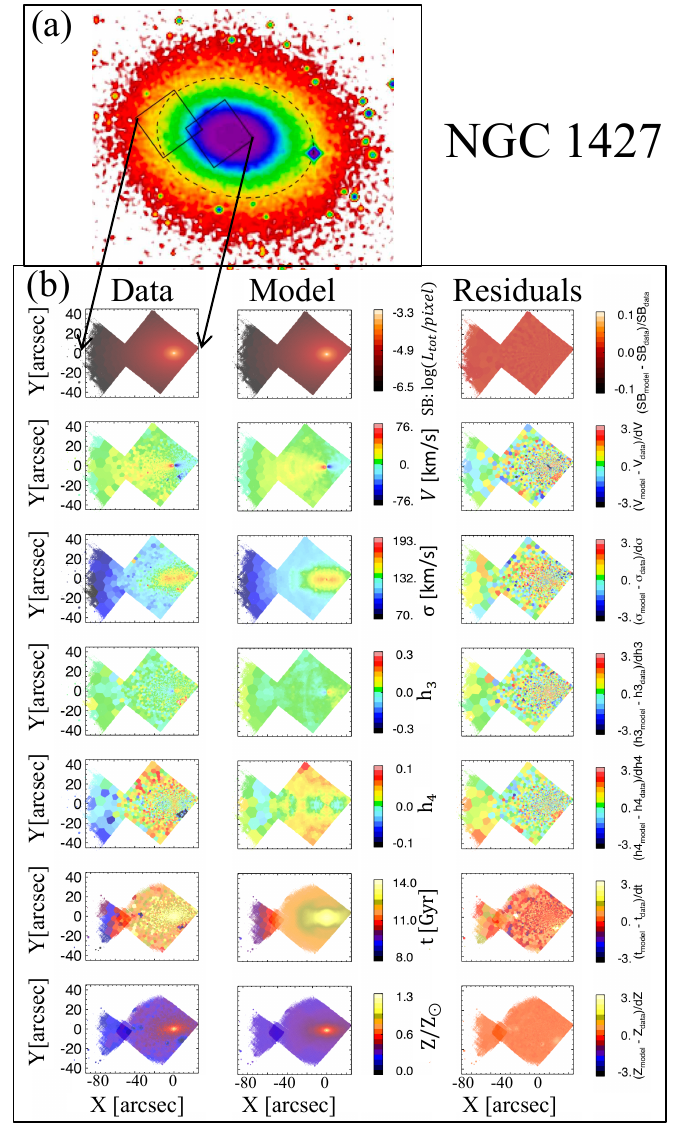} 
\caption{
\textbf{Population-orbit model of NGC\,1427.} See the caption of Figure~\ref{fig:bestfit} for a description.
}
\label{fig:bestfit_1427}
\end{figure*}

For each set of hyper-parameters, we compute tens of thousands of orbits in the corresponding gravitational potential. 
The orbit sampling and calculation follow what we described in \citet{Zhu2020}. 
The weights of the orbits in the resulting orbit library are then determined from fitting simultaneously the orbit-superposition to the intrinsic and projected luminosity density and stellar kinematic maps. 

We adopt a parameter grid with intervals of 0.2, 0.02, 0.02, 0.01 and 0.2 for $M/L_r$, $p$, $q$, $u$, and $\log_{10}(M_{200})$. Then we perform an iterative process searching for the best-fitting models generally following the process described in \citet{Zhu2018a}. As the initial searching steps were already small, we did not refine the search grid around the minimum. At the end, we got 294 models for NGC 1380 and 465 models for NGC 1427 with different collections of the hyper-parameters. 
Once we have explored the hyper-parameter space, we select the best-fitting models with $\Delta\chi^2 \equiv \chi^2- \min(\chi^2) < \sqrt{2 \times n_{\rm GH} \times N_{\rm obs}}$, where $n_\mathrm{GH}=4$ is the number of stellar kinematic moments and $N_\mathrm{obs}$ is the number of bins of each kinematic map. The criterion of $\sqrt{2 \times n_{\rm GH} \times N_{\rm obs}}$ is roughly the $\chi^2$ fluctuation of a single model when perturbing the kinematic data, it generally works well as the $1\sigma$ confidence level according to our test with mock data \citep{Zhu2018a}.

There are 55 models with different values of the hyper-parameters within this $1\sigma$ confidence level for NGC\,1380. We obtain an inclination angle of $77^\circ\pm 2^\circ$, which is consistent with that estimated from the central dust ring \citep{Viaene2019}.
Our dynamical model constrains the stellar mass of NGC\,138 to be $(1.33\pm0.1)\times 10^{11}$ \Msun\, within $2\,R_e$, with $M/L_r = 2.7 \pm 0.3$\,\Msun/\Lsun. 
The mass outside the kinematic data coverage is not well constrained. Therefore, we just take two times of the mass within $R_e$ as NGC 1380's total stellar mass, which gives $\Mstar = (1.8\pm0.2) \times 10^{11}$\,\Msun. 

There are 150 models within this $1\sigma$ confidence level for NGC\,1427.
Our dynamical model gives a total stellar mass of $(4.1\pm0.3)\times 10^{10}$ \Msun\, within $2\,R_e$, with $M/L_r = 2.5 \pm 0.2$\,\Msun/\Lsun, and an inclination angle of $61^\circ\pm 6^\circ$. Also in this case, the mass outside the kinematic data coverage is not well constrained, and so we take two times of the mass within $R_e$ as NGC 1427's total stellar mass, which gives $\Mstar = (5.6\pm0.6) \times 10^{10}$\,\Msun.

%---------------------------------------------------------------------
\subsubsection{Fitting to the age and metallicity maps}
%---------------------------------------------------------------------
Next, for each of the models within the $1\sigma$ confidence level, we applied Voronoi binning in the phase space of $r$ versus $\lambda_z$ to group the orbits into $\sim150$ bundles with comparable weight.
Then we tagged each bundle $k$ with a single stellar age $t_k$ and metallicity $Z_k$ and compared the superposition of the orbit bundles with the light-weighted age and metallicity maps extracted from the VLT/MUSE observations. 

We use Bayesian statistical analysis (Python package pymc3) to estimate the best-fitting age and metallicity for each orbital bundle. To use Bayesian theorem to compute the posterior of a model, we require the prior and the data likelihood. We follow \citet{Zhu2020} in adopting bounded normal and log-normal distributions for the priors on $t_k$ and $Z_k$. While we adopt a student's t-distribution for the data likelihood \citep{Salvatier2016}, which will allow some outliers in the data and results in a robust fitting. For each parameter $t_k$ ($Z_k$), we run a chain with 2000 steps and take the last 500 steps to calculate the mean and $1\sigma$ values of $t_k$ ($Z_k$) for each orbit bundle. Metallicity is fitted separately from age without implementing an age-metallicity correlation. 
The combination of the best-fitting orbital weights with best-fitting age and metallicity of orbital bundles yields population-orbital models that fit in detail the observed stellar kinematic, age, and metallicity maps. 
The best-fitting models of NGC\,1380 and NGC\,1427 are shown in Figures~\ref{fig:bestfit}~and~\ref{fig:bestfit_1427}, respectively.

%---------------------------------------------------------------------
\subsection{Orbit-based decomposition of galaxies}
\label{ss:decomp}
% ---------------------------------------------------------------------

%%%FIG
\begin{figure*}
\centering\includegraphics[width=16cm]{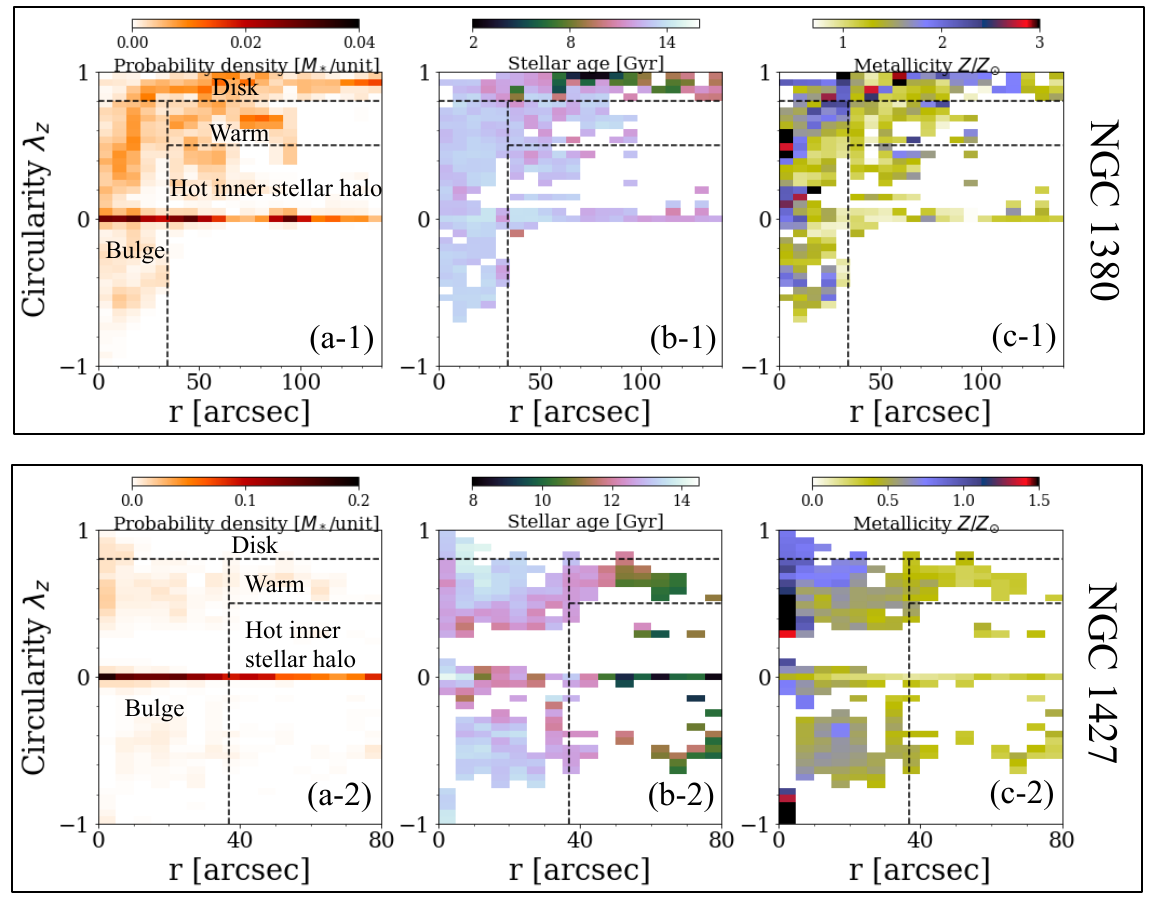}
\caption{
\textbf{Orbital decompositions of NGC\,1380 and NGC\,1427}.
 Panel (a-1): The probability density distribution of the stellar orbits $p(r, \lambda_z)$ in the phase space of time-averaged radius $r$ versus circularity $\lambda_z$ in units of stellar mass per unit area in the phase space. We show the model within 140\,arcsecs (or $\sim14$\,kpc) for NGC\,1380.
 The stellar masses are normalised such that they sum to unity within the data coverage.
 Panels (b-1) and (c-1): The stellar age $t(r, \lambda_z)$ and metallicity $Z(t, \lambda_z)$ distribution of the orbits in the same phase space.  
All distributions $p(r, \lambda_z)$, $t(r,\lambda_z)$, and $Z(t, \lambda_z)$ are averages of multiple best-fitting models that fall within the $1\sigma$ confidence level of the model hyper-parameters.
The dashed lines indicate our orbit-based division into four components: disk, warm, bulge, and hot inner stellar halo. Panels (a-2), (b-2), (c-2): similar for NGC\,1427, which does not have an extended disk.
}
\label{fig:rlz_2gal}
\end{figure*}
%%%FIG

\begin{figure*}
\centering\includegraphics[width=12cm]{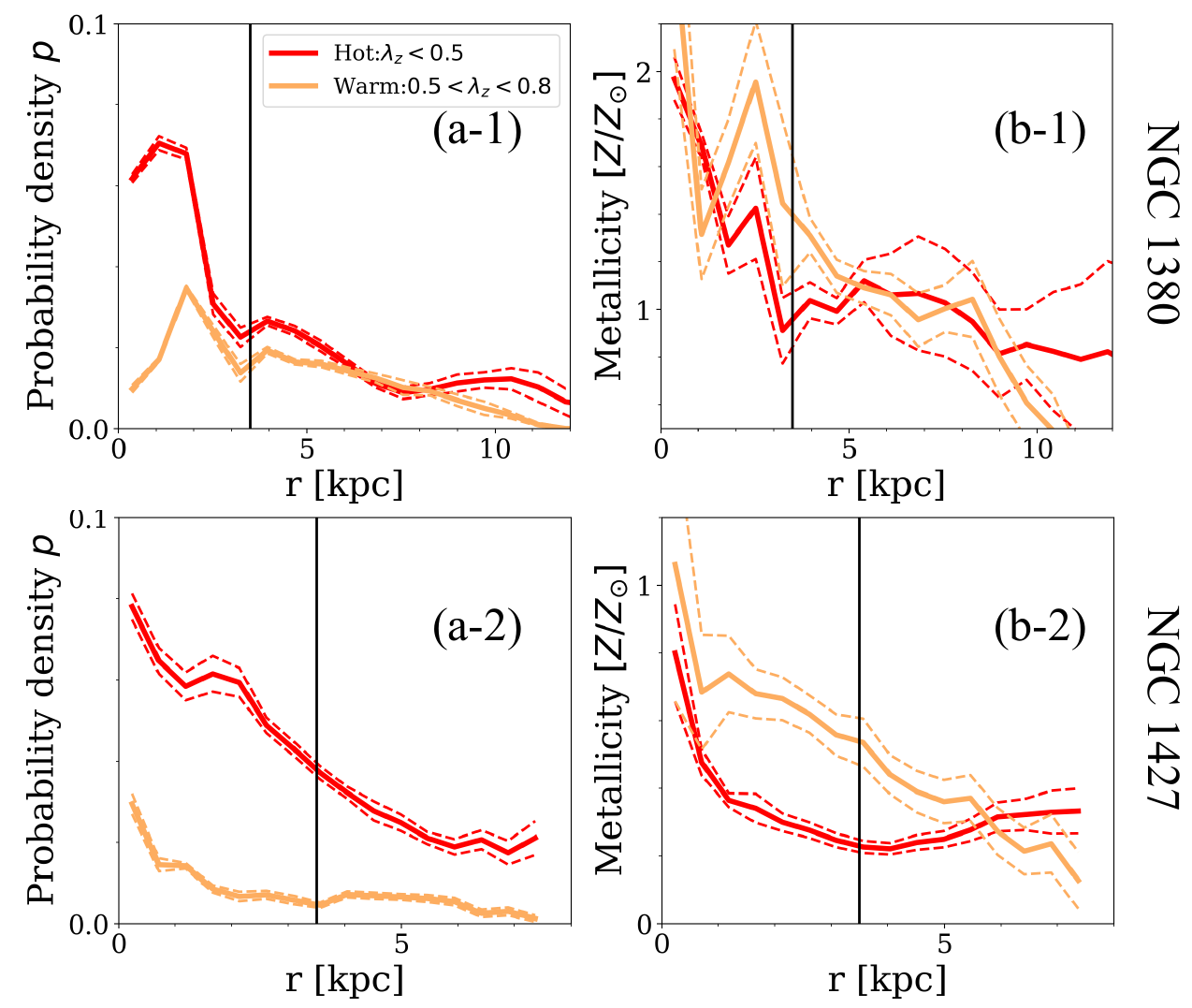}
\caption{{\textbf{Probability density and metallicity profiles.}}
Panel (a-1): Probability density of dynamically hot orbits with $\lambda_z<0.5$ (red) and warm orbits with $0.5<\lambda_z<0.8$ (orange) as a function of radius for NGC\,1380. The red/orange thick and dashed curves are the mean and $1\sigma$ scatter of our best-fitting models. The vertical line at $r=3.5$ kpc marks the separation of bulge and hot inner stellar halo. Panel (b-1): Metallicity distribution of hot and warm orbits as a function of radius for NGC\,1380.  Panels (a-2) and (b-2): similar for NGC\,1427.}
\label{fig:rhot_fcc167}
\end{figure*}

Our method yields the weights of the different stellar orbits that contribute to the best-fitting model as well as the average stellar age and metallicity of the stars represented by each orbit. 
We then characterize each stellar orbit with two main properties \citep{Zhu2018a}: the time-averaged radius, $r$, which represents the extent of the orbit, and the circularity, $\lambda_z = J_z/J_{\rm max} (E)$, which represents the angular momentum of the orbit around the minor $z$-axis normalized by the maximum of a circular orbit with the same binding energy $E$.
Whereas $|\lambda_z| \sim 1$ represents dynamically-cold orbits dominated by regular rotation,  
$\lambda_z \sim 0$ represents dynamically-hot orbits, which are dominated by radial random motions or long-axis tubes with regular rotation around the long axis.
Negative values $\lambda_z<0$ refer to orbits that counter-rotate with respect to the net (prograde) rotation. 
Figure~\ref{fig:rlz_2gal} presents the resulting stellar-orbit distributions of NGC\,1380 and NGC\,1427, respectively, as probability-density distributions $p(r, \lambda_z)$ in the phase space of $r$ versus $\lambda_z$, as well as the age $t(r, \lambda_z)$ and metallicity $Z(r, \lambda_z)$ distributions of the stellar orbits in the same phase space.

Considering the sub-structures in the stellar-orbit probability density and in the age and metallicity distributions in the phase space of $r$ versus $\lambda_z$, we separate each galaxy into four stellar components: 
\begin{itemize}
\item disk ($\lambda_z>0.8$), 
\item bulge ($\lambda_z<0.8$, $r <\rcut$), 
\item warm ($0.5<\lambda_z<0.8$, $\rcut<r<2\,R_e$), 
\item hot inner stellar halo ($\lambda_z < 0.5$, $\rcut<r<2\,R_e$).
\end{itemize}
The latter is in fact a dynamically-hot component, separated from the bulge by $\rcut=3.5$\,kpc, which is generally associated with a transition in density and metallicity distributions of the dynamically-hot orbits.

In Figure~\ref{fig:rhot_fcc167}, we show the probability density and the average metallicity of the dynamical hot ($\lambda_z<0.5$) and warm ($0.5<\lambda_z<0.8$) orbits as a function of radius, for NGC\,1380 (Panel (a-1) and (b-1)) and NGC\,1427 (Panel (a-2) and (b-2)). 
For NGC 1380, both the probability density $p_{\rm hot}$ and metallicity $Z_{\rm hot}$ of the hot orbits
first decreases sharply from an inner peak, and then flatten beyond a transition radius at $\sim3.5$\,kpc.
For NGC 1427, the transition is not so clear in the probability density, but there is a transition in the metallicity profile which flattens beyond $\sim2$\,kpc. A transition generally exists in probability density profile of the hot orbits at $\sim 3.5$ kpc for TNG50 galaxies, but especially those simulated galaxies with high ex-situ fractions do not show a sharp transition similarly to NGC 1427 \citep{Zhu2021}. 
To be able to apply the resulting correlations obtained in \citep{Zhu2021} for simulated galaxies, we adopt the same  $r_{\rm cut} = 3.5$ kpc to separate the bulge and hot inner stellar halo.

The probability density $p_{\rm warm}$ and the average metallicity $Z_{\rm warm}$ of the dynamical warm orbits are also shown for comparison. 
The warm orbits are more metal-rich than hot orbits in NGC 1427, while the difference is small in NGC 1380. In both galaxies, the warm orbits show negative metallicity gradients from the inner to outer regions, which are different from the flattening of metallicity distribution of the hot orbits. 

The four components are defined in such a way as to highlight distinct structural, kinematic, and stellar population distributions, thus likely implying distinct formation pathways. According to previous observational and numerical studies, it would appear that, for galaxies with a quiescent merger history, stars that formed {\it in-situ} from the collapse of dense gas at high redshift constitute a compact spheroid \citep{Du2021,Williams2013}, which constitute a major part of our present-day ``bulge'' and part of the ``warm component'' \citep{Yu2021}. At lower redshift, in-situ stars form out of a gaseous disk that subsequently leads to a dynamical cold stellar disk with strong rotation \citep{Du2021,Yu2021}. No substantial hot inner stellar halo is expected in galaxies with a quiet merger history \citep{Zhu2021}. On the other hand, according to the results of state-of-the-art cosmological galaxy simulations, the hot inner stellar halo is the product of massive mergers \citep{Zhu2021}, and stars in it can have three origins: (1) stars accreted from a massive satellite typically favour highly-radial orbits with $\lambda_z \sim 0$  \citep{Boylan2008, Fattahi2019} and hence partially settle into the bulge and partially stay at larger radii, becoming part of the hot inner stellar halo; (2) at the same time, a merger event can destroy the previously-existing disk and/or bulge of the main progenitor, inducing a fraction of the disk/bulge stars to be kicked to the phase-space regions of the hot inner stellar halo; (3) a merger can last for a few billion years until coalescence and new stars formed during the merger event also partially contribute to the hot inner stellar halo. 

As tested with mock data created from cosmological galaxy simulations, the luminosity fractions of such four components can be recovered in observed galaxies to good accuracy by our modelling (see Appendix~\ref{A:method}).
In the cases of the Fornax cluster galaxies NGC\,1380 and NGC\,1427, the hot inner stellar halo contributes 27 per cent and 34 per cent, respectively, to the total $r$-band luminosity within $2\,R_e$. 
We estimate that the hot inner stellar halo mass is $\Mshalo=(3.6\pm0.5)\times10^{10}$\,\Msun\, and $\Mshalo=(1.4\pm0.2)\times10^{10}$\,\Msun\, for NGC\,1380 and NGC\,1427, respectively. The uncertainty includes contribution of systematically errors evaluated based on the rigorously tests with mock data, and statistical errors of the model, the details are included in the Appendix~\ref{A:uncertainties}.

%---------------------------------------------------------------------
\section{Weighing and timing of past mergers}
\label{S:results}
% ---------------------------------------------------------------------
%%%FIG
\begin{figure*}
\centering\includegraphics[width=15cm]{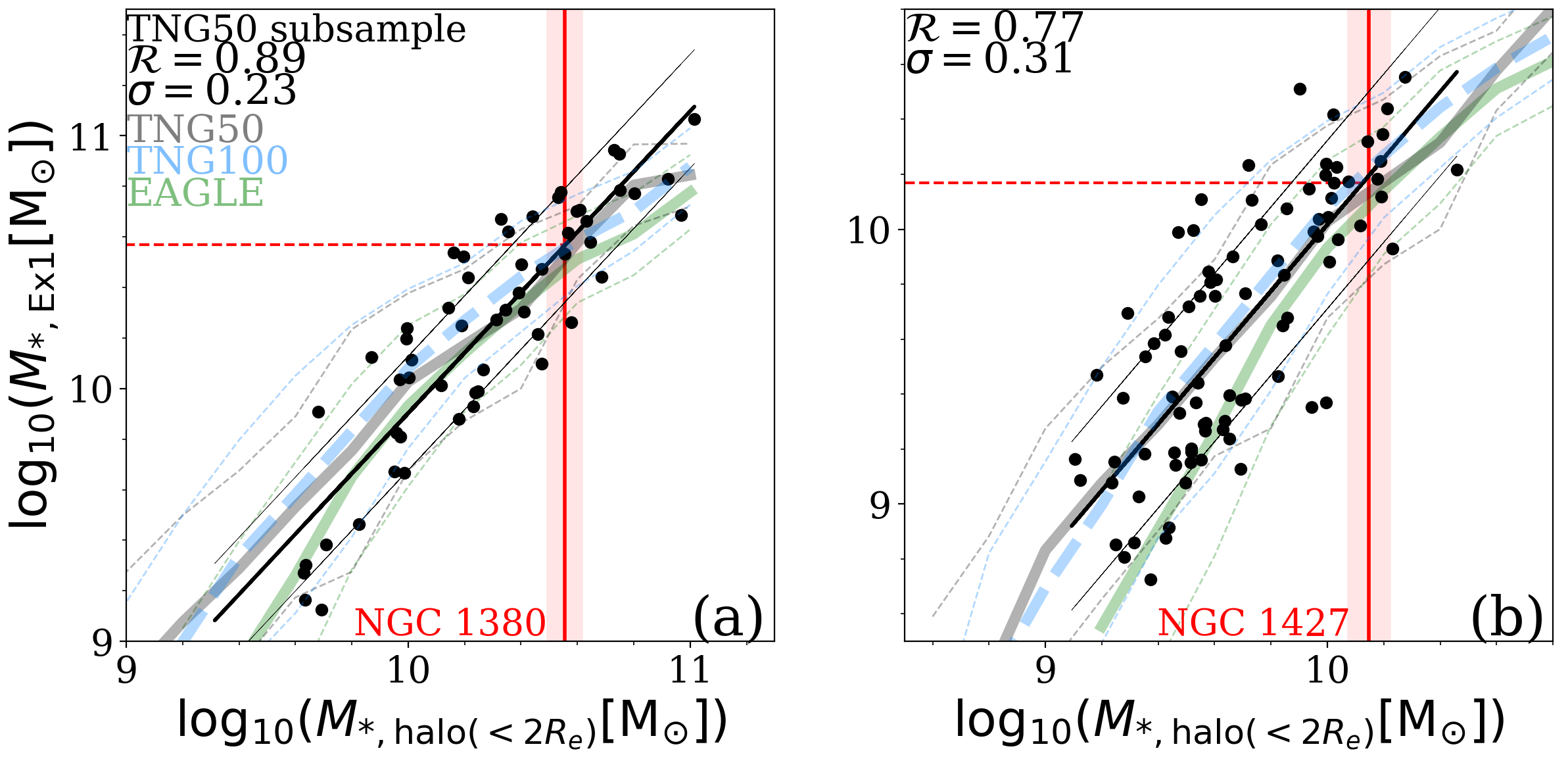}
\caption{
\textbf{Stellar masses of the most massive satellites (now-destroyed) that merged with NGC\,1380 and NGC\,1427.}
Panel (a): The correlation from cosmological simulations between the mass of the hot inner stellar halo \Mshalo\, and the stellar mass $M_{*,\mathrm{Ex1}}$ of the most massive merged satellite. The black dots are a subsample of TNG50 galaxies with similar mass and size as NGC\,1380, the black thick line is a linear fit to the points, while the thin black lines indicate the $\pm1\sigma$ scatter.
The Pearson correlation coefficients $\mathcal{R}$ between $M_{*,\mathrm{halo}(<2R_e)}$ and $M_{*,\mathrm{Ex1}}$ and $1\sigma$ scatter against the linear fit for the TNG50 subsample are indicated. 
The vertical red line with shadows marks the mass and corresponding $\pm1\sigma$ uncertainty of NGC\,1380's hot inner stellar halo. The horizontal dashed line marks the satellite(s) mass inferred from the correlations. The thick and thin curves in grey, blue, and green in the background are the running median and $\pm1\sigma$ scatter of the whole sample of TNG50, TNG100, and EAGLE galaxies.
Panel (b): Similar to Panel (a), but for the subsample of TNG50 galaxies with similar mass and size to NGC1427. The vertical red line marks the mass of NGC\,1427's hot inner stellar halo and the horizontal dashed line marks the satellite(s) mass inferred from the correlations. 
}
\label{fig:mex1}
\end{figure*}
%%%FIG

%%%FIG
\begin{figure*}
\centering\includegraphics[width=15cm]{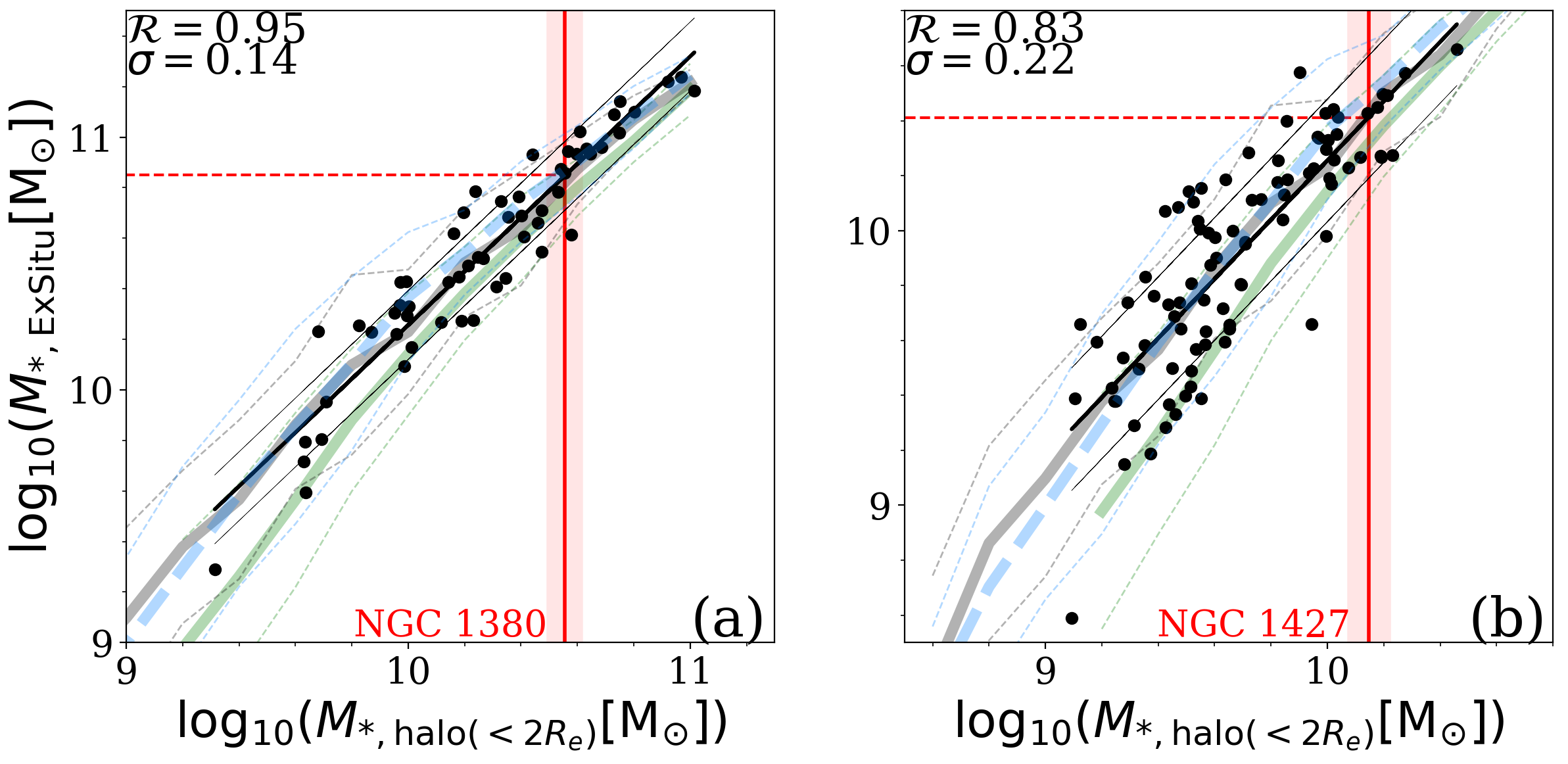}
\caption{
\textbf{Stellar mass accreted from all galaxies ever merged with NGC\,1380 and NGC\,1427, i.e. their total ex-situ mass.}
Panel (a): Similar to Panel (a) of Figure~\ref{fig:mex1}, but Stellar mass $M_{*,\mathrm{ExSitu}}$ accreted from all satellites in the y axis. The black dots are the subsample of TNG50 galaxies with similar mass and size as NGC 1380. The correlation is stronger with larger value of the Pearson correlation coefficients $\mathcal{R}$ and smaller $1\sigma$ scatter than the correlation in Figure~\ref{fig:mex1}. 
The vertical red line with shadows marks the mass and corresponding $\pm1\sigma$ uncertainty of NGC 1380’s hot inner stellar halo. The horizontal dashed line marks the stellar mass accreted from all satellites inferred from the correlations. 
The thick and thin curves in grey, blue and green in the background are the running median and $\pm1\sigma$ scatter of the whole sample of TNG50, TNG100 and EAGLE galaxies. Panel (b): Similar to Panel (a) but for NGC\,1427.
}
\label{fig:mexall}
\end{figure*}
%%%FIG

%---------------------------------------------------------------------
\subsection{Stellar mass of the accreted satellite(s) by NGC\,1380 and NGC\,1427}
%---------------------------------------------------------------------
Cosmological galaxy simulations show that the hot inner stellar halo is a smoking-gun evidence of massive mergers that a galaxy has experienced in its past history. In a companion paper \citep{Zhu2021}, we show that cosmological galaxy simulations return strong correlations between the hot inner stellar halo mass \Mshalo\, of galaxies and the stellar mass from the most massive galaxy they have ever merged with, $M_{*,\mathrm{Ex1}}$ . Combining all the stars from all accreted satellites and mergers, $M_{*,\mathrm{ExSitu}}$, makes the correlations even stronger. These results hold for a definition of the hot inner stellar halo that is identical to the one put forward above and applied to the case of NGC\,1380 and NGC\,1427 via our population-orbit superposition models.
The hot inner stellar halo mass \Mshalo\, can thus be used to infer the mass of the most massive satellite (now destroyed) that a galaxy has ever merged with, as well as of the total accreted or {\it ex-situ} stellar mass of an observed galaxy. This can be done thanks to the fact that the hot inner stellar halo mass can be well recovered by our population-orbit based model for real galaxies, even long after the stars from the merger events are phase-mixed (see Appendix~\ref{A:method}).

%%%FIG
\begin{figure*}
\centering\includegraphics[width=16cm]{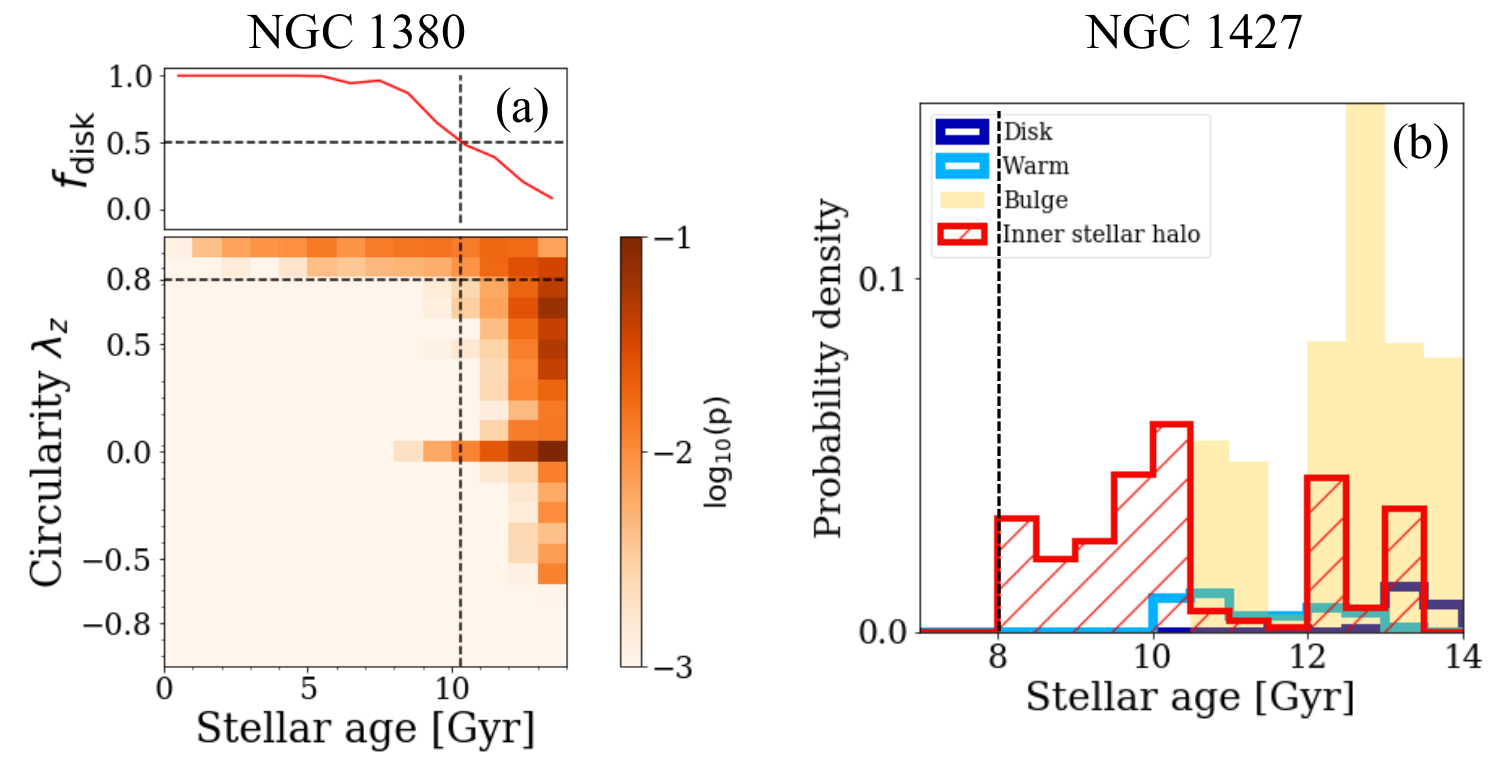}
\caption{\textbf{Timing of the last massive mergers for NGC 1380 and NGC 1427.} 
  Panel (a): The probability distribution of stellar orbits $p(\lambda_z, t)$ in the plane of circularity $\lambda_z$ versus stellar age $t$ from our best-fitting models of NGC\,1380. 
 The fraction of disk orbits $f_\mathrm{disk}$ is plotted in the top subpanel as function of $t$ and shows a sharp transition at $t\simeq10$\,Gyr indicated by the vertical dashed line. We thus infer that the massive merger responsible for the build up of NGC\,1380's hot inner stellar halo ended about $\sim10$\,Gyr ago. Panel (b): The stellar age distribution of disk, warm, bulge, and hot inner stellar halo component from our best-fitting models of NGC 1427. Since the hot inner stellar halo has younger stellar populations than most of the bulge, with youngest stellar population of $\sim8$ Gyr, we infer that the last massive merger responsible for the build-up of hot inner stellar halo could not have ended before that.
}
\label{fig:mergertime}
\end{figure*}
%%% FIG

%---------------------------------------------------------------------
\subsubsection{Stellar mass of the most massive satellite}
\label{ss:Mext1}
%---------------------------------------------------------------------
We infer the stellar mass of the most massive merger for NGC 1380 and NGC 1427 by using the scaling relations predicted by the cosmological simulation TNG50 \citep{Pillepich2019, Nelson2019} (see Figure~\ref{fig:mex1}).
For NGC\,1380 with $M_*= 1.8\times10^{11}$\,\Msun\, and $R_e = 5.8\,$kpc, we compare with the subsample of TNG50 galaxies with $8\times10^{10}<M_*<4\times 10^{11}$\,\Msun\, and $4<R_e<8\,$kpc. 
From these TNG50 galaxies, we obtain $\log_{10}(M_{*,\mathrm{Ex1}}) = 1.195 \times \log_{10}(\Mshalo) -2.04$ with a Pearson correlation coefficient $\mathcal{R}=0.89$, and a $1\sigma$ scatter against of the linear fit of $\sigma = 0.23$. 
With $\Mshalo = (3.6\pm0.5)\times 10^{10}$\,\Msun\, for NGC\,1380, we obtain $M_{*,\mathrm{Ex1}} = 10^{10.57\pm0.24}$\,\Msun\, ($3.7_{-1.5}^{+2.7}\times10^{10}$\,\Msun), with a $3\sigma$ significant lower-limit of $0.7\times10^{10}$\,\Msun. The uncertainty of satellite mass is taken as $d=\sqrt{d_1^2 + d_2^2}$, where $d_1$ is the uncertainty in the derivation of $\log_{10}(\Mshalo)$, while $d_2$ is the uncertainty caused by the scatter of the relation, i.e., $d_2 = \sigma$.

In a similar fashion, for NGC\,1427, with $M_*= 5.6\times10^{10}$\,\Msun\, and $R_e = 4.2\,$kpc, we compare with the subsample of TNG50 galaxies with $4\times10^{10}<M_*<10^{11}$\,\Msun\, and $3<R_e<7\,$kpc. 
From these TNG50 galaxies, we obtained $\log_{10}(M_{*,\mathrm{Ex1}}) = 1.21 \times \log_{10}(\Mshalo) -2.07$ with the Pearson correlation coefficient $\mathcal{R}=0.77$ and $\sigma = 0.31$. With $\Mshalo = (1.4\pm0.2)\times 10^{10}$\,\Msun\, for NGC\,1427, we infer $M_{*,\mathrm{Ex1}} = 10^{10.19\pm0.32}$\,\Msun\ ($1.5_{-0.7}^{+1.6}\times10^{10}$\,\Msun), with a $3\sigma$ significant lower-limit of $0.2\times10^{10}$\,\Msun.

%---------------------------------------------------------------------
\subsubsection{Total accreted stellar masses}
\label{ss:Mextall}
% ---------------------------------------------------------------------
The correlations with \Mshalo\, are stronger when combing all the ever accreted stellar mass $M_{*,\mathrm{ExSitu}}$, as shown in Figure~\ref{fig:mexall}.
The group of galaxies with similar mass and size of NGC\,1380
gives $\log_{10}(M_{*,\mathrm{ExSitu}}) = 1.063  \times \log_{10}(\Mshalo) -0.37$ with $\mathcal{R}=0.95$ and $\sigma = 0.14$.
With $\Mshalo = 3.6\pm0.5\times 10^{10}$\,\Msun\, for NGC\,1380, we obtain $M_{*,\mathrm{ExSitu}} = 10^{10.85\pm0.15}$\,\Msun\, ($7.1_{-2.0}^{+3.0}\times10^{10}$\,\Msun) with a $3\sigma$ significant lower-limit of $2.5\times10^{10}$\,\Msun.

In a similar fashion, the set of galaxies with similar mass and size of NGC\,1427 gives $\log_{10}(M_{*,\mathrm{ExSitu}}) = 1.08 \times \log_{10}(\Mshalo) -0.53$ with $\mathcal{R}=0.83$ and $\sigma = 0.22$. With $\Mshalo = (1.4\pm0.2)\times 10^{10}$\,\Msun\, for NGC\,1427, we infer $M_{*,\mathrm{ExSitu}} = 10^{10.41\pm0.23}$\,\Msun\, ($2.6_{-1.0}^{1.7}\times10^{10}$\,\Msun) with a $3\sigma$ significant lower-limit of $0.5\times10^{10}$\,\Msun.

Our results depend insignificantly, within the reported errorbars, on the particular subset of simulated galaxies adopted for the inference and in fact do not depend either on the specific cosmological galaxy simulation that the relationships have been extracted from. The correlations fitted for the two subsamples of galaxies chosen for NGC\,1380 and NGC\,1427 are consistent with each other. In Figure~\ref{fig:mex1} and Figure~\ref{fig:mexall}, we overlap the correlations \citep{Zhu2021} inferred from all galaxies with $10^{10.3}<M_*<10^{11.6}$\,\Msun\, for three cosmological galaxy simulations whose data are publicly available and that we have analyzed in \cite{Zhu2021}: TNG50 \citep{Pillepich2019, Nelson2019}, TNG100 \citep{Nelson2019release} and EAGLE \citep{McAlpine2016}. The median as well as the $1\sigma$ scatter of these correlations are almost identical in the \Mshalo\, regimes of NGC\,1380 and NGC\,1427: the difference of $M_{*,\mathrm{Ex1}}$ and $M_{*,\mathrm{ExSitu}}$ inferred from these three sets of simulations are smaller than 0.1 dex.

%----------------------------------------------------
\subsection{Timing of the past major mergers}
\label{ss:time}
% ----------------------------------------------------
The time of the mergers in observed galaxies can be inferred from the stellar age distributions in the different dynamical components.
A dynamically-cold disk made of stars with $\lambda_z >0.8$ is usually, at least partially, destroyed during a massive merger, resulting in stars of similar age as the surviving disk but on dynamically-warmer orbits.  
Only after the merger, and in the presence of sufficient gas, new stars can form and re-grow a stellar disk: this structure can persist for a long time if not disrupted by subsequent massive merger events, whereas minor mergers are expected to only heat up the disk without markedly affecting its circularity distribution \citep{Read2008}. This scenario is supported, for example, by the analysis of data in our Milky Way \citep{Belokurov2020}.

NGC\,1380 has a massive dynamically-cold disk component.
We study the distribution of stars in the circularity $\lambda_z$ versus stellar age $t$ plane, $p(\lambda_z, t)$, illustrated by Panel (a) of Figure~\ref{fig:mergertime} for our model of NGC 1380. At any given stellar age $t$, we can calculate the fraction of disk-like orbits $f_{\rm disk}$.
Adopting $f_{\rm disk} = 0.5$ as the transition from spheroid-dominated to disk-dominated regions, the transition time $t|_{f_{\rm disk} = 0.5}$ can be associated with the end of the last massive merger. This transition time $t|_{f_{\rm disk} = 0.5}$ can be well recovered by our model for the galaxies with a massive cold disk at the time of inspection, as is the case for NGC\,1380 (see Appendix~\ref{A:method}).
We find such a transition in our model of NGC 1380, as shown in Figure~\ref{fig:mergertime}: for ages younger than $t\simeq10$\,Gyr, suddenly stars are only on cold orbits with $\lambda_z > 0.8$, i.e., $f_{\rm disk}$ approaches unity, with the transition time of $t|_{f_{\rm disk} = 0.5} = 10$\, Gyr. 
We thus infer that the massive merger responsible for the build up of NGC 1380’s hot inner stellar halo ended $\sim 10$ Gyr ago. Note that the pre-existing cold disk of NGC\,1380 was not fully destroyed by the merger, and the surviving disk mainly locates in the inner regions of the galaxy (see also panel b-1 of Figure~\ref{fig:rlz_2gal}).

NGC 1427 does not have a massive cold disk today, and so possibly it has not re-grown a massive disk after its most massive merger event. Instead, we find that the stars in its hot inner stellar halo are younger than most stars in the bulge; moreover, the hot inner stellar halo holds the youngest stars of the galaxy. According to our understanding of galaxy evolution, see above, inner-halo stars should have formed before the massive merger ended, either in a pre-existing disk of the main progenitor or in the merging companion before and during the merger, because of the triggering of star formation. The youngest stars in the hot inner stellar halo therefore set an upper-limit to the end time of the massive merger. Such a scenario is illustrated by a NGC 1427-like galaxy from TNG50 in the Appendix. From the stellar age distribution of the hot inner stellar halo of NGC 1427 we obtain (see Panel b of Figure~\ref{fig:mergertime}) that the massive merger of NGC 1427 has likely ended $\lesssim 8$ Gyr ago.

NGC\,1380 is located in the so-called north-south clump of the Fornax cluster \citep{Iodice2019b, Nasonova2011}, comprised of a group of galaxies. 
NGC 1427 is the brightest elliptical galaxy on the eastern side of the Fornax cluster. Both galaxies are thought to be accreted into the potential well of the massive cluster core more than 8\,Gyr ago \citep{Iodice2019b}.
The galaxy-galaxy merger frequency in cluster environments is expected to be low because of the high relative velocity of their members \citep{Makino1997,DeLucia2004}: it is therefore expected that most Fornax galaxies underwent their massive mergers before they fell into the Fornax cluster.
The merger of NGC\,1380 with its most massive satellite occurring $\sim10$\,Gyr ago is consistent with this scenario. Two other early-type galaxies in the same clump, NGC 1380A (FCC 177) and NGC 1381 (FCC 170), have also been suggested to have accreted a substantial fraction of stars $\sim 9$\,Gyr ago \citep{Pinna2019, Pinna2019b, Poci2021}. 
On the other hand, for NGC 1427 and considering that the merger process could last for $1-2$ Gyr, the constraint from the stellar age distribution of its hot inner stellar halo does allow the inferred merger event to have started before infall.  
Whereas we cannot set a strong lower-limit on the time of the ending of the most massive merger of NGC 1427, we notice that mergers in a cluster may still happen if two galaxies fell into the cluster within the same group \citep{Tran2005,Delahaye2017,watson2019}.

\begin{figure*}
\centering\includegraphics[width=15cm]{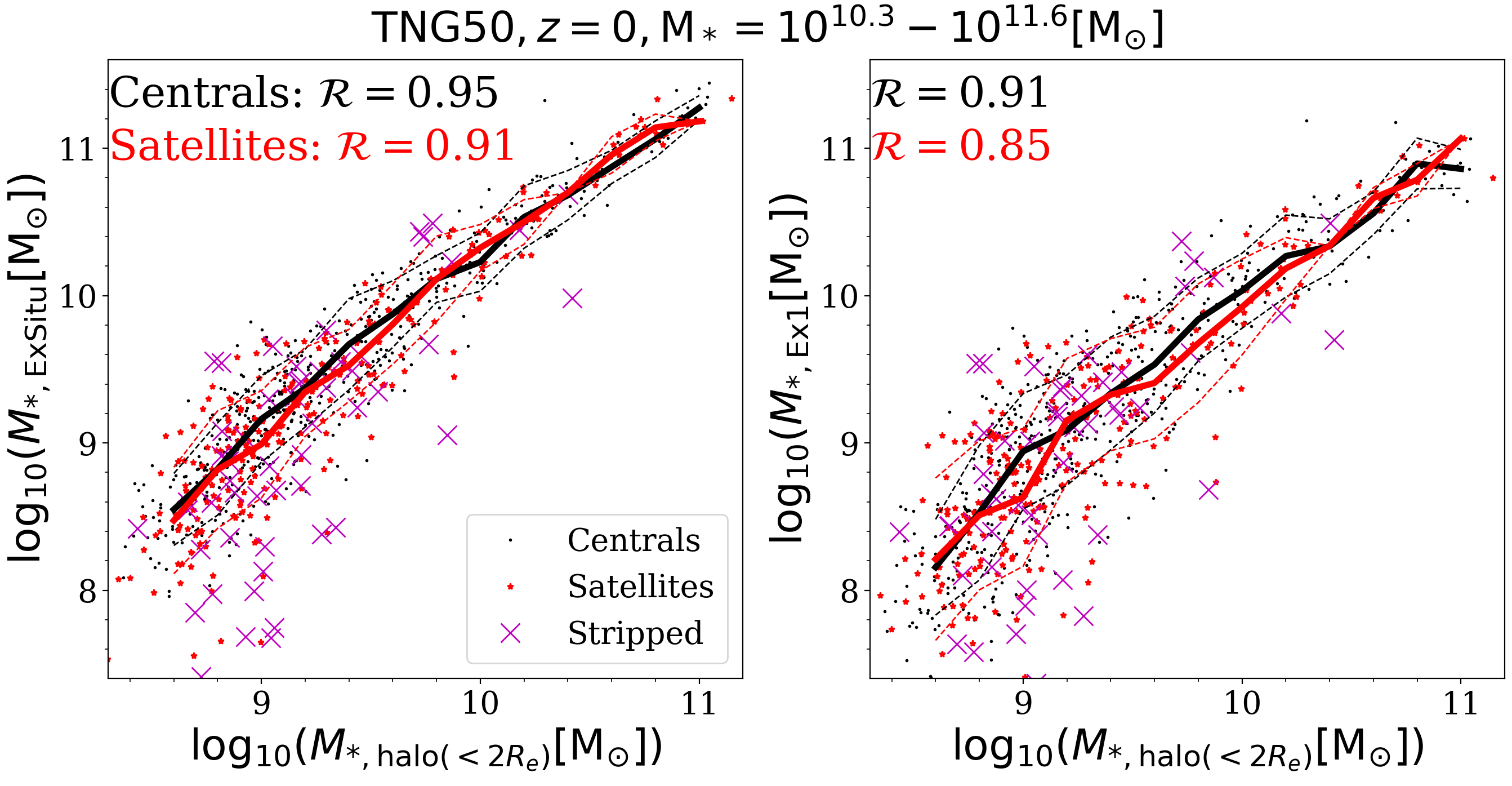}
\caption{
\textbf{Correlations of hot inner stellar halo mass \Mshalo\, vs. accreted stellar mass $M_{*,\mathrm{Ex1}}$ (left) and $M_{*,\mathrm{ExSitu}}$ (right) for central and satellite galaxies separately.} All TNG50 galaxies with $M_*=10^{10.3}-10^{11.6}$\,\Msun\, are included. The black dots are central galaxies and the red dots are satellites. The thick black and red curves are the running median for central and satellite galaxies, respectively, while the thin dashed curves are the corresponding $\pm1\sigma$ scatter. The satellites with more than $20\%$ of stellar mass stripped are marked by magenta crosses. The correlations are almost the same for central and satellite galaxies. The strongly stripped satellites tend to locate below the median of the correlations, some of them deviate from the correlations significantly below the lower '$-1\sigma$' curves.}
\label{fig:mex_strip}
\end{figure*}

\begin{figure*}
\centering\includegraphics[width=15cm]{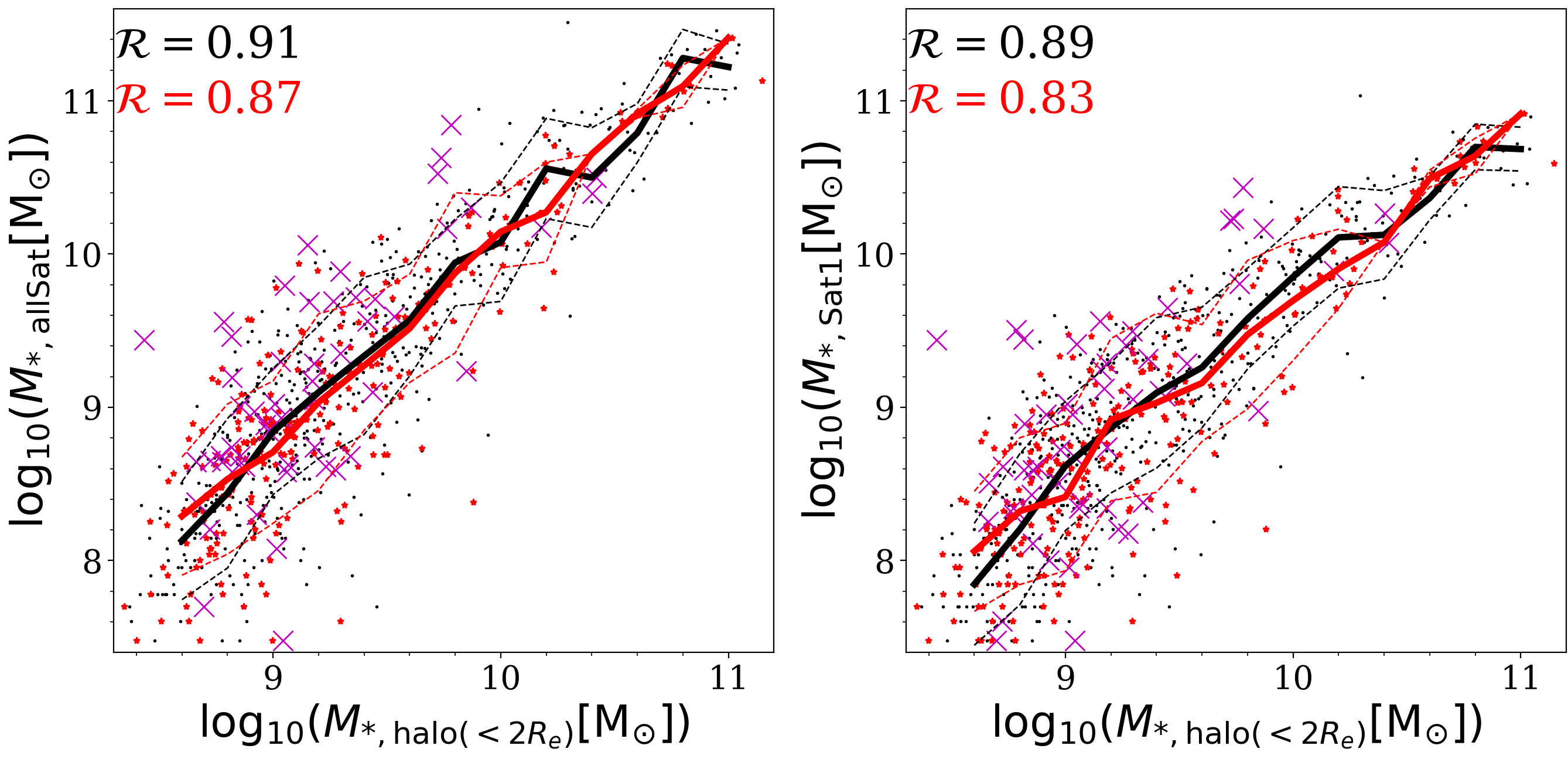}
\caption{
\textbf{The correlations by using the secondary progenitor mass $M_{*,\rm allSat}$ and $M_{*, \rm Sat1}$ instead of the ex-situ stellar mass in present-day galaxies $M_{*,\rm ExSitu}$ and $M_{*, \rm Ex1}$.} Symbols are the same as in Figure~\ref{fig:mex_strip}. The strongly stripped galaxies move upward, some of them become outliers above the upper '$+1\sigma$' curves. 
}
\label{fig:msat1}
\end{figure*}

%----------------------------------------------------
\section{Effects of cluster environment}
\label{S:discussion}
% ----------------------------------------------------
%---------------------------------------------------------------------
Since both NGC\,1380 and NGC\,1427 are in the Fornax cluster, their internal structures might be altered by physical processes like `harassment’ or `tidal shocking' in the cluster gravitational potential \citep[e.g.,][]{Joshi2020}. TNG50 covers a large variety of group/cluster environments, including ten clusters within the mass range of $M_{\rm 200}\sim 10^{13.3}-10^{14.3}$\,\Msun, similar to the Fornax cluster. We can hence directly check with TNG50 whether, and by how much, high-density environments affect the inner stellar halos of satellite galaxies.

In particular, we inspected all the satellite galaxies with $M_*>10^9$\,\Msun\, in the second most massive halo of TNG50, which has $M_{\rm 200} = 9.4\times 10^{13}$\,\Msun. By comparing the stellar orbit distribution of each galaxy at the time it fell into the cluster to its distribution at $z=0$, we find that for $\sim 80\%$ of the galaxies, the hot inner stellar halo mass \Mshalo\ remains almost unchanged, with differences smaller than 0.1 dex. For the remaining $\sim20\%$ of the galaxies, with relatively low mass and that fell into the cluster early, \Mshalo\ decreases significantly as these galaxies get strongly stripped by the cluster. There are a few galaxies whose \Mshalo\ slightly increases, mostly within 0.1 dex. For these systems, their pre-existing cold disks were heated up, possibly by harassment, with most of the stars becoming dynamically warm and a small fraction of them becoming hot. There are no satellite galaxies with a significantly hot inner stellar halo component induced after falling into the cluster, due to the fact that no massive mergers happen for these galaxies after falling into the cluster. The massive hot inner stellar halos of the two galaxies studied in this work are thus unlikely to be induced by either `harassment’ or `tidal shocking' in the cluster environment.

To further illustrate the possible effects of cluster environments on our results, we show the correlations of the hot inner stellar halo mass \Mshalo\, vs. accreted stellar masses $M_{*,\mathrm{Ex1}}$ and $M_{*,\mathrm{ExSitu}}$ for the central and satellite galaxies from TNG50 separately. We use all TNG50 galaxies with $M_* = 10^{10.3}-10^{11.6}$\,\Msun, except 37 with ongoing mergers -- the same sample as used in \citet{Zhu2021}. We quantify the stripped mass of a galaxy by comparing its present-day stellar mass $M_{*, \rm z=0}$ with the maximum stellar mass it has ever reached $M_{*, \rm max}$. 
All satellite galaxies with $M_{*,\rm z=0} < 0.8 \; M_{*, \rm max}$ are considered to be strongly stripped. 
In Figure~\ref{fig:mex_strip}, we can see that the correlations are almost the same for central and satellite galaxies except the strongly stripped ones. Most of the strongly stripped galaxies have $\Mshalo<10^{10}$\,\Msun, they tend to be located below the median of the correlations, and some of them are located significantly below the lower '$-1\sigma$' curves.

Instead of $M_{*,\mathrm{Ex1}}$ and $M_{*,\mathrm{ExSitu}}$, we can quantify a merger by the secondary progenitor mass defined as the maximum of the stellar mass it has ever reached before the merger ended \citep{Rodriguez2016}. 
We use $M_{*,\mathrm{Sat1}}$ to indicate the stellar mass of the most massive secondary progenitor a galaxy has ever accreted, and $M_{*,\mathrm{allSat}}$ to indicate the sum of all the secondary progenitors more massive than $1\%$ stellar mass of the current galaxy. $M_{*,\mathrm{Ex1}}$ is highly correlated and larger than $M_{*,\mathrm{Sat1}}$ by a factor of $\sim30\%$ for all the central galaxies and for all the satellite galaxies that are not strongly stripped\footnote{A merger could last for 1-2 Gyr during which star formation and disruption can continue. Thus stars accreted from the secondary progenitor are more than it has had at any moment.}. On the other hand, for the strongly stripped galaxies, $M_{*,\mathrm{Ex1}}$ calculated from the current galaxy are significantly smaller than the original accreted stellar mass, and also smaller than $M_{*,\mathrm{Sat1}}$.
In Figure~\ref{fig:msat1}, we show the correlations by using $M_{*,\rm allSat}$ and $M_{*, \rm Sat1}$ instead of $M_{*,\mathrm{ExSitu}}$ and $M_{*,\mathrm{Ex1}}$. The strongly stripped galaxies, which were located below the $-1\sigma$ curves in Figure~\ref{fig:mex_strip}, move upward, and tend to follow the relations in Figure~\ref{fig:msat1}. Some of the strongly stripped galaxies become outliers above the upper '$+1\sigma$' curves. 

In summary, for all central and most satellite galaxies, $M_{*,\mathrm{Ex1}}$ and $M_{*,\mathrm{Sat1}}$ are equally valid to quantify their past merger events, while \Mshalo\, defined in the current galaxy is a good indicator for the merger mass. However, if a satellite galaxy is strongly stripped, then both $M_{*,\mathrm{Ex1}}$ and \Mshalo\, can be affected, and the former more than the latter. The accreted stellar mass $M_{*,\mathrm{Ex1}}$ inferred from the median correlation in Figure~\ref{fig:mex_strip} could be higher than $M_{*,\mathrm{Ex1}}$ it has in the current galaxy, but lower than the original $M_{*,\mathrm{Ex1}}$ it once had. Similarly, $M_{*,\mathrm{Sat1}}$ inferred from the median correlation in Figure~\ref{fig:msat1} is likely to be under-estimated if the hot inner stellar halo was strongly stripped.

For our two galaxies in the Fornax cluster, NGC 1380 is a massive lenticular galaxy with $M_* \sim 1.8\times10^{11}\,M_{\odot}$, with median size for its stellar mass, and with an extended disk. Its outer isophotoes at $R>3$ arcmin appear twisted with respect to the inner more disky regions \citep{Iodice2019a}, which could be the result of the major merger. All these features suggest that NGC\,1380 is unlikely to be strongly stripped and the results presented above should not be affected by environmental effects.
NGC\,1427 is a massive elliptical galaxy with $M_* \sim 5\times10^{10}\,M_{\odot}$, also with median size for its stellar mass, and with a massive hot inner stellar halos with $\Mshalo>10^{10}\,M_{\odot}$. The outer isophotes of NGC\,1427 appear asymmetric, suggesting it could be partly stripped. However, its inner regions -- and thus also the hot inner stellar halo -- are unlikely to be significantly affected. We expect that \Mshalo\, is still a good indicator for its past merger mass, although the ex-situ stellar mass in the current galaxy might actually be lower.

\begin{table*}
\caption{NGC\,1380's and NGC\,1427's discovered major merger events in comparison with past mergers in the Milky Way and the Andromeda galaxy. 
}
\scriptsize\centering
\label{tab:mwm31}
\begin{tabular}{*{6}{l}}
\hline
       Galaxy name & $M_*$ &  $M_{*,\rm Ex1}$ &   Time of Merger & Redshift &  $M_{*,\rm ExSitu}$\\
\hline
Milky Way & $ 5 \times10^{10}$\,\Msun  &$3-6\times 10^8$\,\Msun  & $ 10$ Gyr ago & $z\sim1.8$  & -\\
Andromeda & $1.3\times10^{11}$\,\Msun  &$2\times 10^{10}$\,\Msun & $ 2$ Gyr ago &  $z\sim0.15$  &-\\
NGC 1380 & $(1.8\pm0.2)\times10^{11}$\,\Msun &$3.7_{-1.5}^{+2.7}\times 10^{10}$ \,\Msun & $10$ Gyr ago & $z\sim1.8$  &$7.1_{-2.0}^{+3.0}\times10^{10}$\,\Msun \\
NGC 1427 & $(5.6\pm0.6)\times10^{10}$\,\Msun   & $1.5_{-0.7}^{+1.6}\times 10^{10}$\,\Msun & $t\lesssim 8$ Gyr ago & $z\lesssim 1$  & $2.6_{-1.0}^{+1.7}\times10^{10}$\,\Msun\\
  \hline
\hline
 \end{tabular}
 \tablefoot{The six columns from left to right are: galaxy name, galaxy's current total stellar mass $M_*$, stellar mass of the most massive merger $M_{*,\rm Ex1}$, time of the last massive merger, redshift of the last massive merger, total ex-situ stellar mass $M_{*,\rm ExSitu}$. The total stellar mass of the Milky Way is adopted from \citet{Bland-Hawthorn2016}, while its merger mass and time are from \citet{Helmi2018} and \citet{Belokurov2020}. The total stellar mass of
 Andromeda is adopted from \citet{Corbelli2010}, while its merger mass and merger time are from \citet{DSouza2018b}. The errors on NGC 1380 and NGC 1427 indicate the $1\sigma$ uncertainties of our results.}
\end{table*}

%----------------------------------------------------
\section{Summary}
\label{S:summary}
% ----------------------------------------------------

We discovered that the two early-type galaxies NGC\,1380 and NGC\,1427 in the Fornax cluster both underwent an ancient massive merger, and constrained the merger mass and time. 

By applying our recently-developed population-orbit superposition method to the observed luminosity density, stellar kinematic, age and metallicity maps, we obtained the stellar orbits, age and metallicity distribution of each galaxy. We then decomposed each galaxy into four components, which included a dynamically hot inner stellar halo as a relic of past massive mergers.
By comparing to analogues from cosmological galaxy simulations, chiefly TNG50, we used the robustly measured mass of the hot inner stellar halo to infer the stellar mass from the ever-accreted most massive satellite $M_{*,\rm{Ex1}}$, as well as the total ex-situ stellar mass $M_{*,\rm{ExSitu}}$ for these two galaxies.
We further found that the last merger of NGC\,1380 ended $\sim 10$ Gyr ago based on the stellar age distribution of its re-grown dynamically cold disk, in a way comparable to how the last massive merger time of the Milky Way could be inferred \citep{Belokurov2020}. 
The merger in NGC 1427 ended at $t\lesssim 8$ Gyr ago based on the stellar populations in its hot inner stellar halo, in a way comparable to how the merger time in M31 was estimated \citep{DSouza2018b}.
While the quantification of the satellite masses and merger times in both the Milky Way and M31 rely on observations of individual stars, we have overcome these limitations by applying our new population-orbit superposition method to IFU data. 

In Table~\ref{tab:mwm31}, we list the mass and time of the most massive merger events of NGC\,1380 and NGC\,1427 in comparison with those of the Milky Way and M31. The Milky Way, M31 and NGC 1380 are the only three galaxies for which both the merger mass and time have been quantitatively uncovered so far.
The mass of the most massive merger of NGC\,1427 is comparable to that of M31, while the mass of the most massive merger of NGC\,1380 is twice as large.
However, the mergers in NGC\,1380 and NGC\,1427 happened much earlier than in M31.
Given the subsequent evolution in the cluster environment, the high encounter velocities and long relaxation times have resulted in smoother structural distributions in NGC\,1380 and NGC\,1427 than in M31.
The most massive merger in the Milky Way is similarly old, but is smaller with stellar mass of $3-6\times 10^8$ \Msun, which is only about $\sim1.5$ per cent of NGC 1380’s and $\sim3$ per cent of NGC 1427's progenitor satellite stellar mass. 
The merger event in NGC\,1380 is the oldest and most massive one uncovered in nearby galaxies so far.

Similar IFU data as we used for NGC\,1380 and NGC\,1427 could be obtained for hundreds of nearby galaxies through observations with instruments like MUSE.
Our discovery thus opens a new way to quantify the merger history of nearby galaxies and so place important constraints on the assembly of galaxies as a function of environment and in a cosmological context.

\begin{acknowledgement}
We are grateful for discussions with Illustris TNG team members Dylan Nelson, Lars Hernquist, Ruediger Pakmor and Mark Vogelsberger.
LZ acknowledges the support from the National Key R$\&$D Program of China under grant No. 2018YFA0404501 and National Natural Science Foundation of China under grant No. Y945271001. 
GvdV acknowledges funding from the European Research Council (ERC) under the European Union's Horizon 2020 research and innovation programme under grant agreement No 724857 (Consolidator Grant ArcheoDyn). 
EMC is supported by MIUR grant PRIN 2017 20173ML3WW$\_$001 and Padua University grants DOR1885254/18, DOR1935272/19, and DOR2013080/20. 
FP, IMN, and JFB acknowledge support through the RAVET project by the grant PID2019-107427GB-C32 from the Spanish Ministry of Science, Innovation and Universities (MCIU), and through the IAC project TRACES which is partially supported through the state budget and the regional budget of the Consejer\'ia de Econom\'ia, Industria, Comercio y Conocimiento of the Canary Islands Autonomous Community. 
TNG50 was realised with compute time granted by the Gauss Centre for Super-computing (GCS), under the GCS Large-Scale Project GCS-DWAR (2016; PIs Nelson/Pillepich).
\end{acknowledgement}

%---------------------------------------------------------------------
\bibliographystyle{aa}  % style aa.bst
\bibliography{ms_fcc167} % your references Yourfile.bib
%---------------------------------------------------------------------

%---------------------------------------------------------------------
\appendix
%---------------------------------------------------------------------

%---------------------------------------------------------------------
\section{Method validation}
\label{A:method}
% ---------------------------------------------------------------------
We previously tested the ability of our population-orbit method to recover the age and metallicity properties of different orbital components \citep{Zhu2020}. However, there we did so only for simulated spiral galaxies that do not have without a significant hot inner stellar halo within the data coverage. In this Appendix, we extend such tests. 

The major degeneracy that our model could present is between the disk and the halo, as a face-on disk might look similar to a halo in the kinematic maps.
NGC\,1427 has a small disk in the inner regions with $r<10\arcsec$. The disk fraction in the outer regions is very small. Therefore, the disk-halo degeneracy should not be a problem for NGC\,1427.
For NGC\,1380, on the other hand, we have performed a series of tests using mock data from four simulated galaxies from TNG50: TNG50\,468590, TNG50\,375073, TNG50\,2, TNG50\,220596, which are chosen to have similar hot inner stellar halo mass and disk mass as NGC\,1380.

The mock data are created in a similar way as done before \citep{Zhu2020}, but restricted to have the same inclination angle $\vartheta=77^\circ$, similar data coverage and data quality as for NGC\,1380. We first project a simulated galaxies into the observational plane with inclination angle $\vartheta=77^o$, place it at a distance of $d = 25$ Mpc, and observe it with a pixel size of 1\arcsec. Then we sum up the mass of the stellar particles within each pixel to obtain a surface mass density map (i.e., a mock image of the galaxy), by assuming that light equals mass.

We choose the area with $-40\arcsec<x<40$\arcsec\, and $-35\arcsec<y<35$\arcsec\, to create the kinematic maps. This data coverage is chosen to be similar to the observed kinematic maps of NGC\,1380, where $x$ is along the major axis and $y$ is along the minor axis of the galaxy.
According to the number of stellar particles in each pixel, we perform a Voronoi binning process to reach a target $(S/N)_T=50$, assuming Poission noise $\sim \sqrt{N_{\rm particles}}$. Given all the particles in each Voronoi bin, we derive the mass-weighted mean velocity, velocity dispersion, $h_3$, and $h_4$ by fitting a GH function \citep{vanderMarel&Franx1993} to the stellar line-of-sight velocity distribution (LOSVD), as well as we calculate mass-weighted average age ($t$) and metallicity ($Z/Z_{\rm sun}$). Note that here the mean velocity and velocity dispersion are not directly calculated from the LOSVD, but they are actually the parameters of the base Gaussian in the GH function. 

After the Voronoi binning, the spatial resolution of our mock data is $\sim 150-1000$ pc, which is comparable to that of the kinematic data (binned with target $(S/N)_T=100$) from the Fornax 3D project \citep{Sarzi2018}. 
We use a simple function inferred from the CALIFA data to construct the errors for kinematic maps \citep{Tsatsi2015}, with the errors proportional to $\frac{(S/N)_T}{(S/N)_{\rm bin}} (1 + 1.4\log{N_{\rm pix}})$,  where $(S/N)_{\rm bin} =  \sqrt{N_{\rm particles}}$, and $N_{\rm pix}$ is the number of pixels in a Voronoi bin. We have larger errors for kinematics in the area with lower surface density, where more pixels are included in each bin in order to reach the target $(S/N)_T$. For age and metallicity, the tests on the full-spectrum fitting to mock spectra of $(S/N)_T=40$ return random errors of $10\%$ \citep{Pinna2019}. The errors could be lower for spectra with higher $S/N$, while it could be higher for real spectra due to possible systematic effects. As proof-of-concept, we adopt relative errors of $10\%$ for age and metallicity. So, to include the noise, the kinematics, age and metallicity maps are perturbed by random numbers, normally distributed with dispersions equal to the observational errors. We find that the error maps of the mock data are similar to the data of MUSE observations for galaxies in the Fornax 3D project \citep{Sarzi2018}.

\begin{figure*}
\centering\includegraphics[width=10cm]{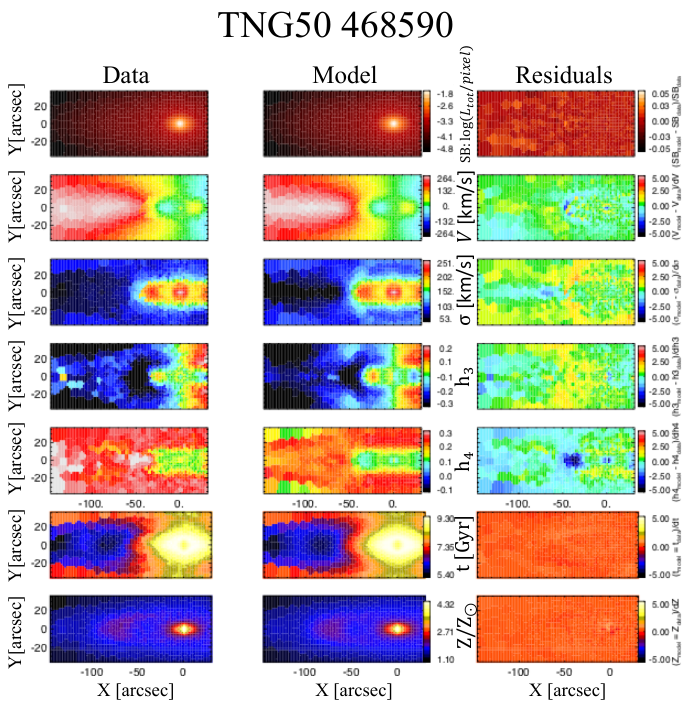}
\caption{
\textbf{Population-orbit model of the simulated galaxy TNG50\,468590.}
The first column shows the mock data with, from top to bottom, surface mass density, mean velocity, velocity dispersion, Gauss-Hermite coefficients $h_3$, $h_4$, mass-weighted age and metallicity maps. The second column is the best-fitting population-orbit model, and the third column are the residuals of data minus model, divided by the errors assigned assuming similar quality data as for NGC\,1380.
}
\label{fig:bestfit_tng}
\end{figure*}

We take each mock data set as a real observed galaxy and apply the population-orbit superposition model to it. 
In Figure~\ref{fig:bestfit_tng}, we show the mock data created from TNG50\,468590 and our best-fitting population-orbit model to it. The model fits well the data in great detail.

%%%FIG

%---------------------------------------------------------------------
\subsection{Recovery of the four components.}
%---------------------------------------------------------------------
\begin{table*}
\caption{Model recovery of the four components. For each of the four simulated galaxies used for testing our methods, we list the mass fraction, average age and metallicity of the disk, bulge, warm component, and hot inner stellar halo within a $2\,R_e$ radius. The first values are the ground truth calculated from the simulations directly; the values in the brackets are the mean and standard deviation of our best-fitting models within $1\sigma$ confidence level.
}
\scriptsize\centering
\label{tab:error}
\begin{tabular}{*{5}{l}}
\hline
               & Disk  &  Warm &  Bulge &  hot inner stellar halo     \\
  \hline
  TNG50 468590 & & & & \\
  Luminosity fraction & $0.208(0.188\pm0.004)$  & $0.042(0.066\pm0.004)$ & $0.418(0.395\pm0.004)$ &  $0.333(0.351\pm0.004)$ \\
  Stellar age [Gyr]   & $4.8(5.1\pm0.1)$        & $9.2(7.5\pm0.4)$       &  $9.0 (9.0\pm0.04)$    & $9.4(9.1\pm0.2)$ \\
  Metallicity [$Z/Z_{\odot}$] & $2.1(2.4\pm0.1)$ & $1.3 (1.4\pm0.1)$     &  $3.1(2.9\pm0.1)$     & $1.46(1.51\pm0.05)$ \\
 
  \hline
  TNG50 375073 & & & & \\
  Luminosity fraction & $0.241(0.203\pm0.003)$  &  $0.028 (0.054\pm0.002)$ &  $0.509(0.515\pm0.003)$ & $0.221(0.228\pm0.005)$ \\
  Stellar age [Gyr]   & $4.7(4.2\pm0.2)$        & $8.8(8.5\pm0.9)$       &  $9.3 (9.3\pm0.1)$    & $9.7(9.9\pm0.2)$ \\
  Metallicity [$Z/Z_{\odot}$] & $1.9(2.4\pm0.1)$ & $1.3 (1.5\pm0.1)$     &  $3.0(2.9\pm0.1)$     & $1.41(1.39\pm0.04)$ \\
 
\hline
  TNG50 220596 & & & & \\
  Luminosity fraction & $0.151(0.122\pm0.003)$  &  $0.158 (0.149\pm0.006)$ &  $0.477(0.484\pm0.004)$ & $0.214(0.244\pm0.008)$ \\
  Stellar age [Gyr]   & $8.3(6.9\pm0.3)$        & $9.1(8.1\pm0.3)$       &  $10.4 (10.4\pm0.1)$    & $9.7(10.5\pm0.2)$ \\
  Metallicity [$Z/Z_{\odot}$] & $2.7(3.1\pm0.2)$ & $2.0 (2.0\pm0.1)$     &  $2.8(2.7\pm0.1)$     & $1.69(1.75\pm0.05)$ \\
 
 \hline
  TNG50 2 & & & & \\
  Luminosity fraction & $0.136(0.107\pm0.004)$  &  $0.15 (0.17\pm0.01)$ &  $0.416(0.402\pm0.004)$ & $0.30(0.31\pm0.01)$ \\
  Stellar age [Gyr]   & $6.0(5.6\pm0.2)$        & $6.7(6.4\pm0.3)$       &  $9.3 (9.2\pm0.1)$    & $9.0(9.1\pm0.2)$ \\
  Metallicity [$Z/Z_{\odot}$] & $2.2(2.5\pm0.2)$ & $1.9 (2.1\pm0.1)$     &  $2.7(2.6\pm0.1)$     & $1.70(1.67\pm0.07)$ \\
  
  \hline
\hline
 
 \end{tabular}
\end{table*}

%%%FIG
\begin{figure*}
\centering\includegraphics[width=16cm]{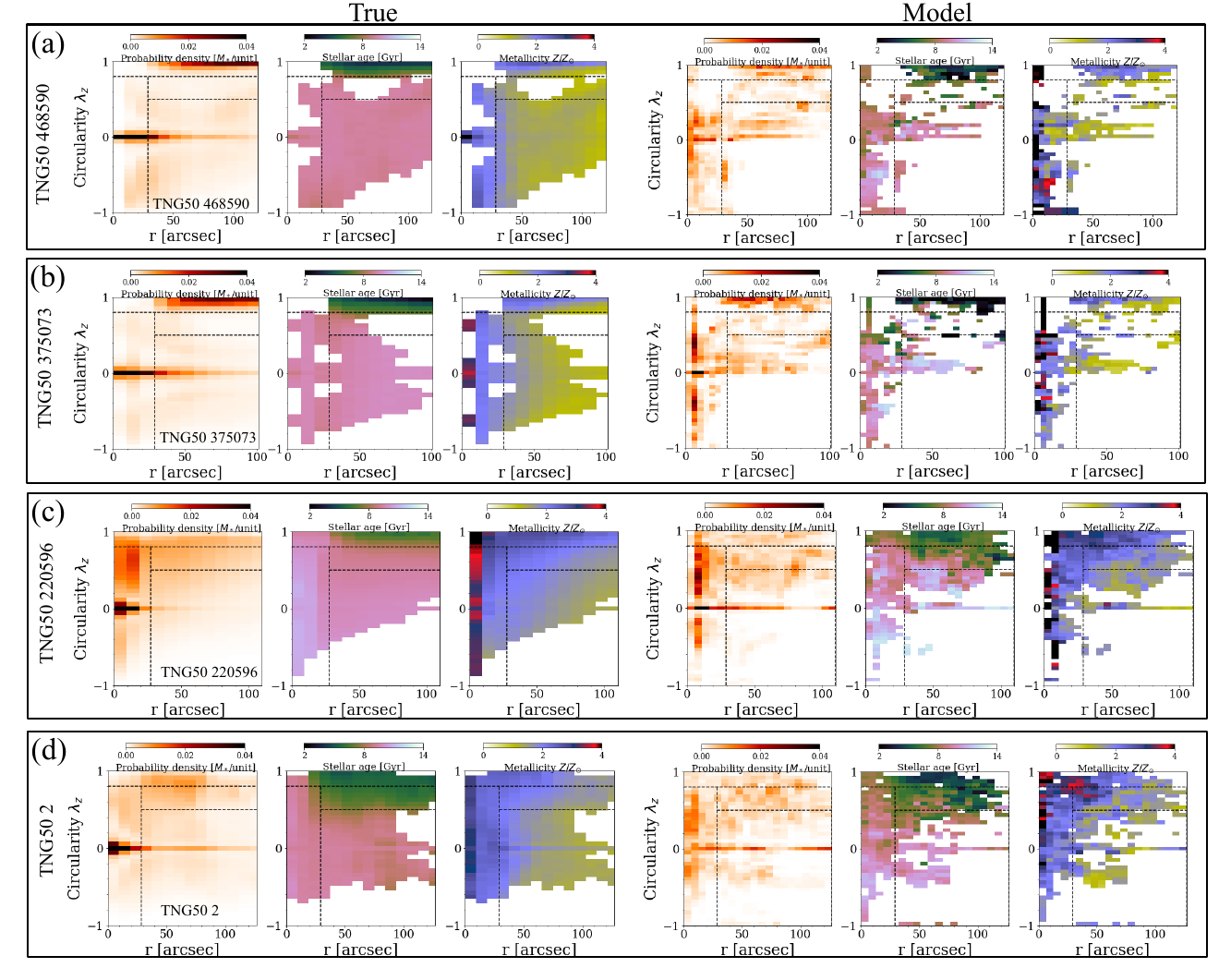}
\caption{
  \textbf{Verification of the population-orbit method with mock data created from galaxies of the TNG50 simulation:} TNG50\,468590, TNG50\,375073, TNG50\,220596, TNG50\,2 from Panel (a) to Panel (d).
For each galaxy, the left three plots show the distributions directly from the simulation, which are well matched by those extracted from the best-fitting population-orbit models in the right three plots.
From left to right, we show the distributions in phase space of the time-averaged radius $r$ versus circularity $\lambda_z$ of orbital probability density $p(r, \lambda_z)$, age $t(r, \lambda_z)$, and metallicity $Z(\lambda_z, r)$, all are shown only within $2\,R_e$. 
The dashed lines indicate our orbital-based division into four components: disk, bulge, warm component, and hot inner stellar halo.
The corresponding mass fractions within a $2\,R_e$ radius are listed in Table~\ref{tab:error}.}
\label{fig:test_rlz4}
\end{figure*}
%%%FIG

%%%FIG
\begin{figure*}
\centering\includegraphics[width=16cm]{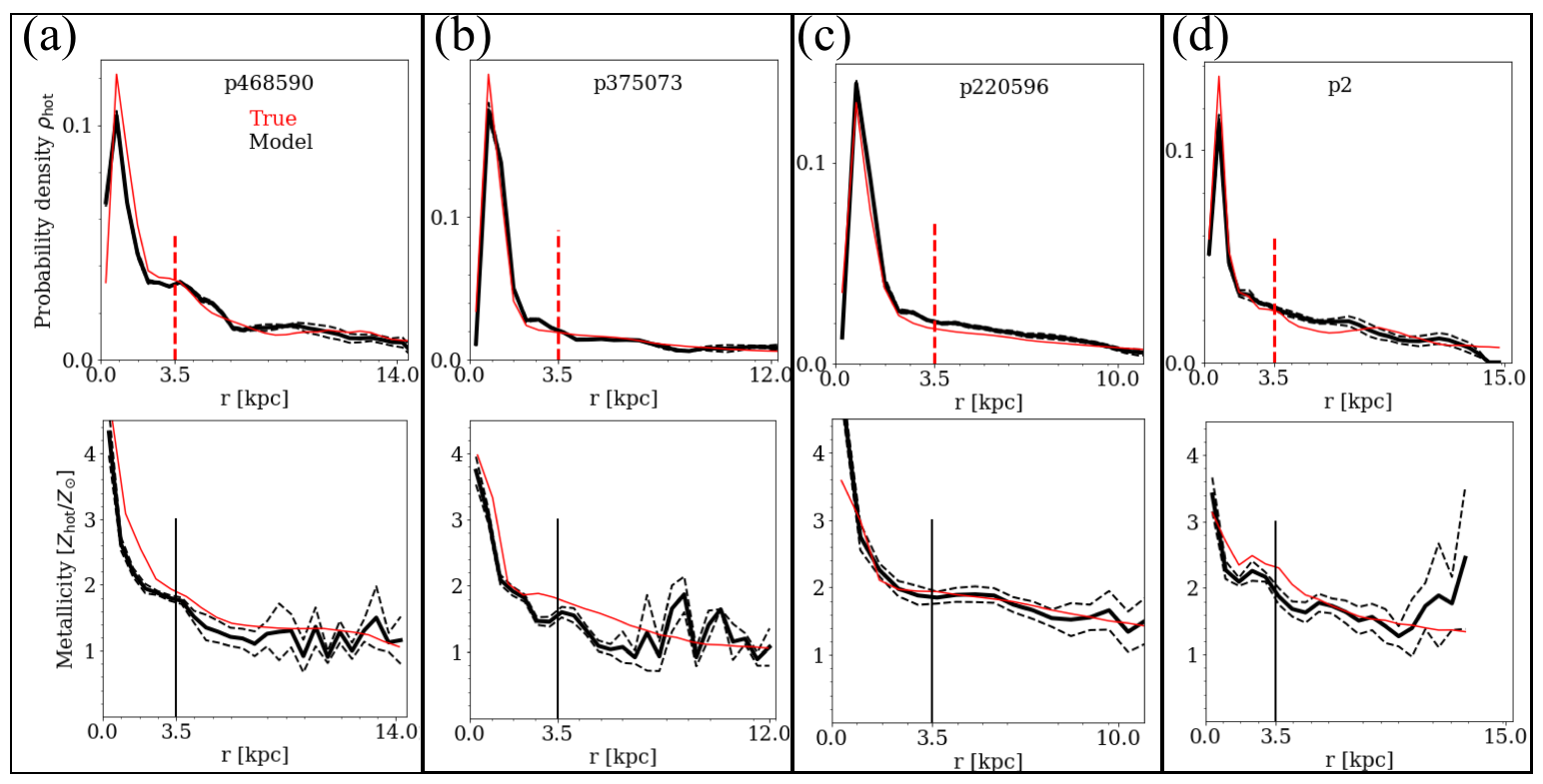}
\caption{\textbf{Verification of the bulge versus hot inner stellar halo decomposition.}
Probability density $p_{\rm hot}$ (top panel) and metallicity $Z_{\rm hot}/Z_{\odot}$ (bottom panel) of the hot orbits as a function of radius for TNG50\,468590, TNG50\,375073, TNG50\,220596, and TNG50\,2 from Panel (a) to Panel (d). 
In each panel, the red curve is the true profile from the simulation, the black thick and dashed curves are the mean and $1\sigma$ scatter of our best-fitting models. The vertical lines mark the position of $r=3.5$ kpc where we separate the bulge from hot inner stellar halo. }
\label{fig:test_pd4}
\end{figure*}
%%%FIG

\begin{figure*}
\centering\includegraphics[width=16cm]{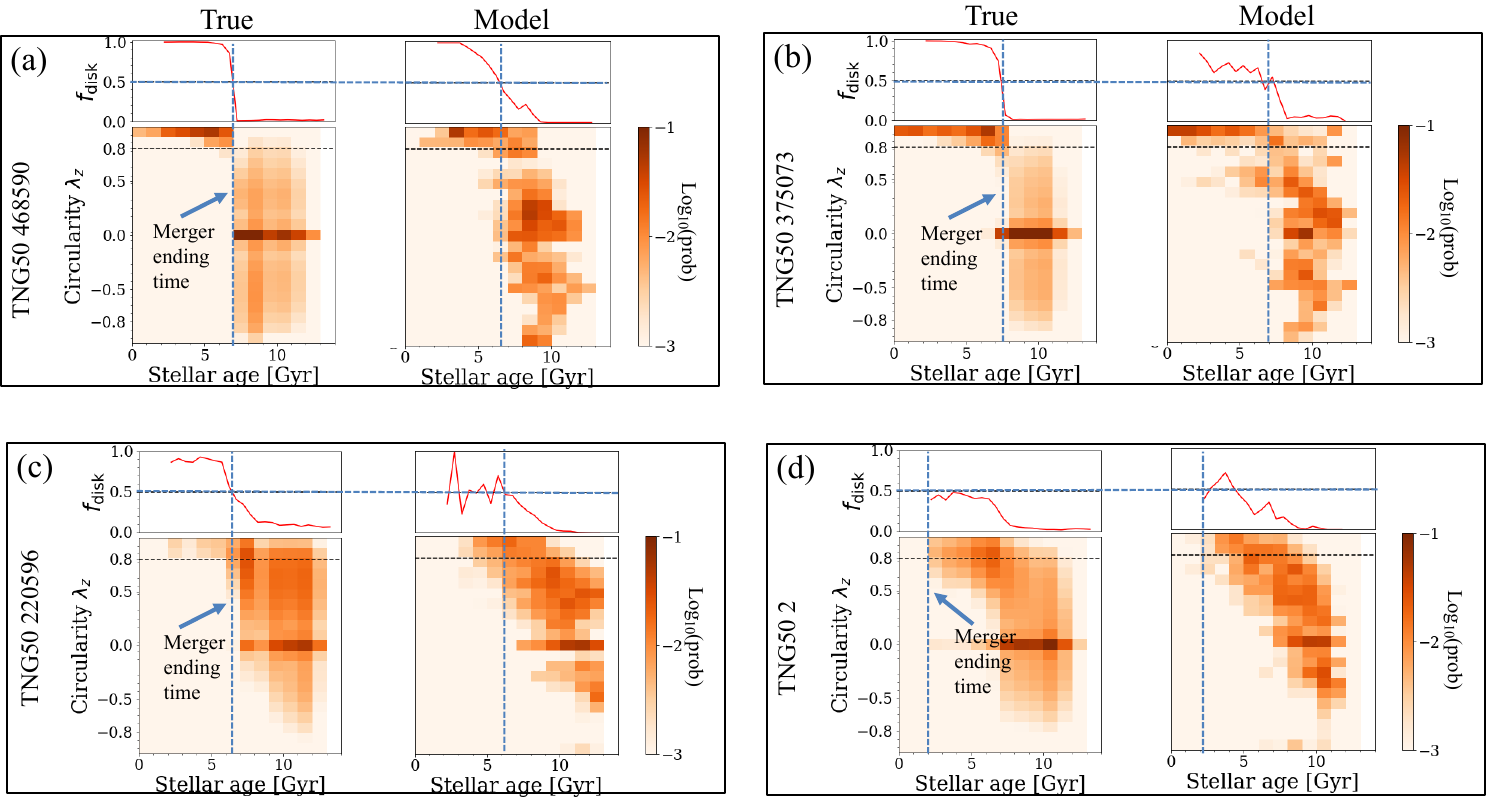}
\caption{
  \textbf{Verification of the method to time mergers from the circularity-age distribution.} Probability density distributions in phase space of stellar age $t$ versus circularity $\lambda_z$ for TNG50\,468590, TNG50\,375073, TNG50\,220596, and TNG50\,2 from Panel (a) to Panel (d).
  In each panel, the left plot is the true distribution for the stellar orbits within $2\,R_e$ directly obtained from the simulation, the vertical dashed line marks the merger end time of the last massive merger the galaxy has experienced. The merger end time is consistent with the time $t|_{f_{\rm disk}=0.5}$ in TNG50 468590, TNG50 375073, TNG50 220596, with still significant star formation after the last massive merger, and it is consistent with the end of star formation in TNG50 2. The right plot is the distribution from our best-fitting model, and the vertical dashed line marks the time $t|_{f_{\rm disk}=0.5}$ in the model for TNG50 468590, TNG50 375073, TNG50 220596, while it marks the end of star formation in TNG50,2.
 }
\label{fig:test_tlz4}
\end{figure*}
%%%FIG

\begin{figure*}
\centering\includegraphics[width=16cm]{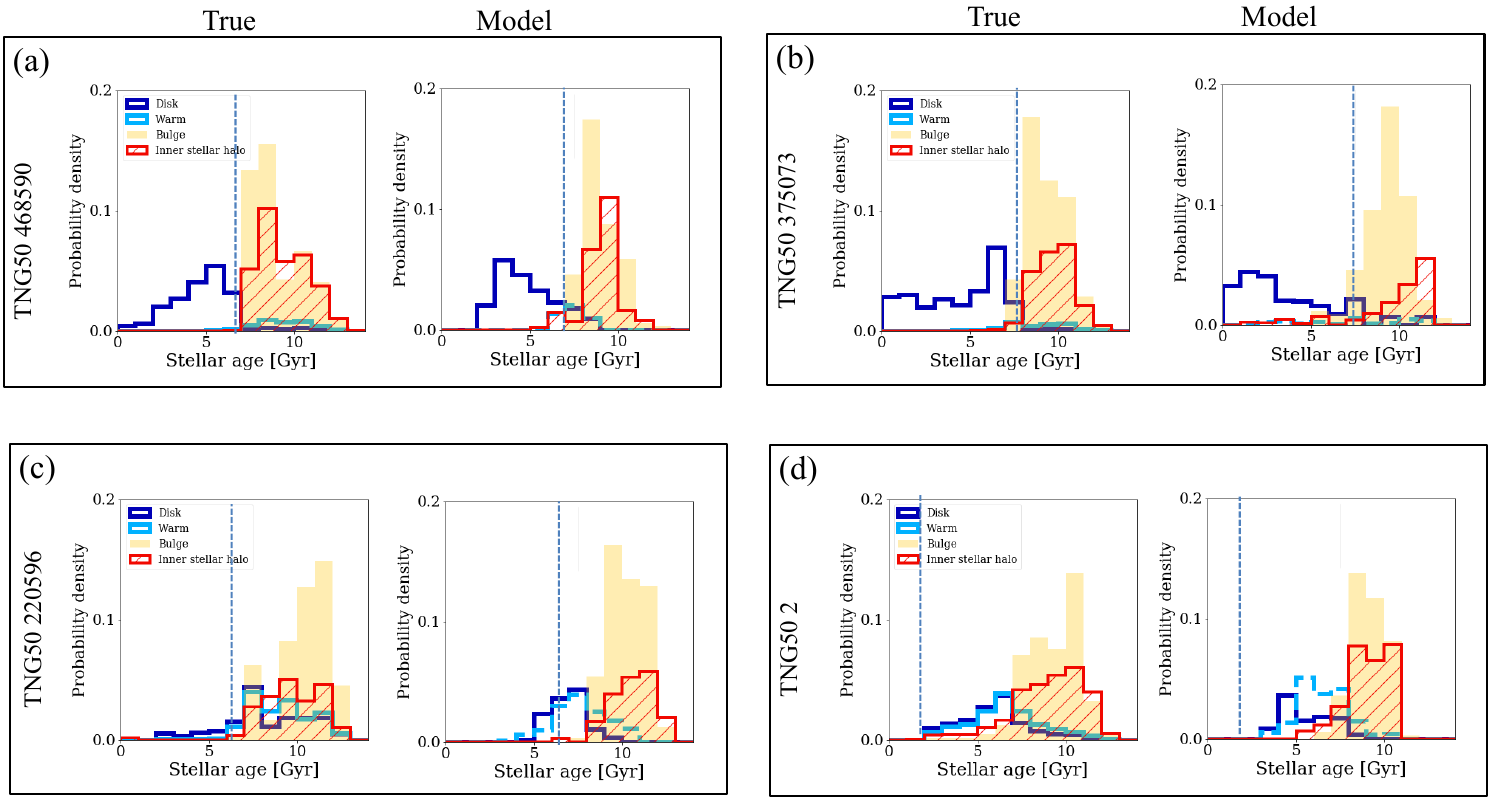}
\caption{
  \textbf{Verification of the method to assign a time limit to the merger from the age distributions of the galaxy components.}
 Stellar age distributions of the disk, warm, bulge and hot inner stellar halo components for TNG50\,468590, TNG50\,375073,  TNG50\,220596, and TNG50\,2 from Panel (a) to Panel (d). In each panel, the left plot is the true stellar age distribution from the stellar particles in the simulation, the right plot is the distribution from our best-fitting model, and the vertical dashed line marks the end of the last massive merger.
 }
\label{fig:test_tdis4}
\end{figure*}

In Figure~\ref{fig:test_rlz4}, we show that the probability density distribution $p(\lambda_z, r)$, age distribution $t(\lambda_z, r)$, and metallicity distribution $Z(\lambda_z, r)$ directly measured from the simulations are well matched by those extracted from the best-fitting population-orbit models.
There is a discrepancy in the inner parts, where the bulge in the simulated galaxies is dominated by box orbits with $\lambda_z = 0$ whereas part of the box orbits in the model are represented actually by tube orbits with non-zero but opposite values of $\lambda_z$. 
The latter, however, does not affect our division into four components:
disk ($\lambda_z>0.8$), 
bulge ($\lambda_z<0.8$, $r <\rcut$), 
warm component ($0.5<\lambda_z<0.8$, $r >\rcut$), 
and hot inner stellar halo ($\lambda_z < 0.5$, $\rcut<r <2R_e$), where $\rcut = 3.5$ kpc is chosen for all cases.
The mass fraction, average age and metallicity of the disk, bulge, warm component, and hot inner stellar halo within a $2\,R_e$ radius for the four simulated galaxies are given in Table~\ref{tab:error}.
Our model excellently recovers the mass fraction, age and metallicity of the hot inner stellar halo, even though the statistical errors obtained from the $1\sigma$ scatter of the best-fitting models are too small to cover the true values in some cases. The absolute difference between the true values and our model recovered mean values are still small. 

In Figure~\ref{fig:test_pd4}, we display the probability density distribution $p_{\rm hot}$ and average metallicity $Z_{\rm hot}$ as a function of radius $r$ by
summing all the hot orbits with $\lambda_z < 0.5$. The red curves are directly calculated from the simulations, while the thick and dashed black curves are the mean and $1\sigma$ scatter from our best-fitting models. The model generally matches the radial profiles for both the probability density and metallicity.
Both $p_{\rm hot}$ and $Z_{\rm hot}$ sharply decrease from an inner peak, and flatten at larger radii.

From the above analysis, we find that, although our data have limited extension along the galaxy minor axis, we still are able to uncover the intrinsic stellar orbit distributions quite well. This is thanks to the high data quality that provide reliable high-order GH coefficients $h_3$ and $h_4$, which in turn encode the information of the combination of disk and hot inner stellar halo. The age and metallicity of stellar orbits in the hot inner stellar halo from our model still have relatively large fluctuations. Having data extending more along the galaxy minor axis should improve the constraints on the stellar age and metallicity distribution of the hot inner stellar halo.

%---------------------------------------------------------------------
\subsection{Recovery of the merger's end time.}
%---------------------------------------------------------------------
%%%%%
We further check how well we can recover the correlation between stellar age $t$ and circularity $\lambda_z$.
In Figure~\ref{fig:test_tlz4}, we show the probability density of orbits in the circularity $\lambda_z$ versus stellar age $t$, $p(\lambda_z, t)$, of the four simulated galaxies elected for the testing (panels a-d). In each panel, the left plot gives the true distribution from the simulation and the right plot shows our best-fitting model. We know every aspect of the merger history of the simulated galaxies: the real end time of the last massive merger of each simulated galaxy is marked in Figure~\ref{fig:test_tlz4}. In three of the four galaxies, i.e., TNG50 468590, TNG50 375073, TNG50 220596, there is substantial star formation after the last massive merger. Stars formed after the last massive mergers rarely ended on non-disk orbits. In all these cases, a sharp transition of stars from non-disk dominated orbits to disk dominated orbits can be clearly identified: therefore we can confidently associate this transition to the last massive merger end time.
The time of $t|_{f_{\rm disk}=0.5}$ is a good proxy for the merger end time.

In practise, $t|_{f_{\rm disk}=0.5}$ is well recovered by our model in TNG50\,468590 and TNG50\,375073, which have massive cold disks formed after the last massive merger, although it is not as sharp as in the simulation data due to uncertainties of the stellar age of orbits in our model. In TNG50\,220596, only a small fraction of disk formed after the merger: even though our model generally matches the distribution $p(\lambda_z, t)$, the transition time $t|_{f_{\rm disk}=0.5}$ is hard to recover. 
In TNG50\,2, there is no disk formed after the last merger: throughout the stellar age range, the cold disk never becomes dominating. The merger end time is consistent with the star formation end time. Our model generally recovers the distribution $p(\lambda_z, t)$, and the star formation end time.
In both TNG50\,220596 and TNG50\,2, a massive merger occurring at low redshift partially heated up the previous cold disk and produced a significant warm component. The stars in such a warm component exhibit a wide range of stellar ages, similar to that of the remnant cold orbits (Figure~\ref{fig:test_tlz4}). However they are on average younger than the stars on dynamical hot orbits with $\lambda_z \sim 0$. 
While NGC\,1380 has a massive cold disk, which is younger than any other component, the warm component in NGC 1380 is as old as the stars on hot orbits, like those in TNG50 468590 and 375073. 

In Figure~\ref{fig:test_tdis4}, we analyse the recovery of age distribution of the four components. In the left plot of each galaxy, we show the stellar age distribution of particles belonging to each of the four components in the simulation. In the right plot, we show the stellar age distribution of orbital bundles belonging to each component from our best-fitting models. Our model is able to roughly recover the true age distribution of each component. 

In these four simulated galaxies, inner-halo stars are old and disk stars are young: in some cases, e.g., TNG50 468590 and TNG50 375093, our model introduces some young stars in the inner-halo due to the degeneracy with disk. Similarly for NGC\,1380, we get a small fraction of inner-halo stars with $t<10$ Gyr, which are likely induced by such degeneracy. NGC\,1380 is thus still consistent with the scenario that the inner-halo stars are formed before the merger end time of $10$ Gyr.

In these four simulated galaxies, disk is the youngest component, the youngest stars in the disk being 0, 0, 2, and 2 Gyr old in TNG50 468590, TNG50 375073, TNG50 220596, TNG50 2, respectively: from the best-fitting model, we obtain 2, 0, 4, and 3 Gyr. The declination of probability density toward younger ages is sharper in our model than in the real distribution from the simulation. Similarly for the hot inner stellar halo in NGC\,1427, which is the youngest component in that galaxy. The star formation end time of 8 Gyr obtained by our model is likely an upper limit, and the real star formation end time could be 1-2 Gyr younger. So the merger end time of $t\lesssim 8$ Gyr is a relaxed upper limit.

%---------------------------------------------------------------------
\section{Uncertainties on the stellar mass of the hot inner stellar halo}
\label{A:uncertainties}
%---------------------------------------------------------------------

We have shown that the mass fraction of the hot inner stellar halo is well recovered by our population-orbit superposition model fitted to the mock data of NGC\,1380-like simulated galaxies from TNG50. The mass fractions of the hot inner stellar halo $f_{\rm halo}$ of the four TNG50 test galaxies is 0.333, 0.221, 0.214, and 0.30, respectively: we get $0.351\pm0.004$, $0.228\pm0.005$, $0.244\pm0.008$ and $0.31\pm0.01$ from the best-fitting models. Here the error is the statistical error taken $1\sigma$ scatter of the best-fitting models. The statistical error is very small, and the true value is not always covered within the $1\sigma$ statistical error. As the absolute difference between the true and model-estimated values is 0.018, 0.007, 0.01, and 0.03 for the four simulated galaxies, we consider the average of these four values as a systematic error, which gives 0.016.
For NGC\,1380, we obtained the luminosity fraction of hot inner stellar halo $f_{\rm halo}$ within $2R_e$ as $0.267\pm0.006$: considering a systematic error of 0.016, we have $f_{\rm halo}=0.267\pm0.017$. The error of 0.017 is determined by $d=\sqrt{d_1^2 + d_2^2}$, where $d_1 = 0.006$ is the statistical error and $d_2 = 0.016$ is the systematic error.
For NGC\,1427, we obtained the luminosity fraction of hot inner stellar halo within $2R_e$ as $0.34\pm0.02$, and considering a systematic error of 0.016, we have $f_{\rm halo} = 0.34\pm0.03$.

Whereas orbit probabilities in the simulated galaxies are mass-weighted, those for NGC\,1380 and NGC\,1427 are $r$-band luminosity weighted. For NGC\,1380, under the simplest assumption of best-fitting constant stellar mass-to-light ratio $\Mstar/L_r = 2.7\pm0.3$ \MLsun\, from our dynamical models, we estimate $\Mshalo = L_r(r<2R_e) \times \Mstar/L_r \times f_{\rm halo} = (3.6\pm0.5)\times 10^{10}$\,\Msun. The uncertainty is calculated by the error propagation formula considering the uncertainties in $\Mstar/L_r$ and $f_{\rm halo}$. Similarly for NGC\,1427, we arrive at $\Mshalo = (1.4\pm0.2)\times 10^{10}$\,\Msun\, with $\Mstar/L_r = 2.5\pm0.2$ given by our dynamical model.
\Mshalo\, will be used throughout to infer the masses of the mergers. 

In NGC\,1380, we know from a previous study on stellar population synthesis that $M_*/L_r$ varies from $4.8\pm 0.7 $ \MLsun\, in the bulge-dominated inner regions to $3.5\pm0.5$ \MLsun\, in the inner-stellar-halo-dominated outer regions by considering a variable IMF \citep{Navarro2019}. The adopted error is the $1\sigma$ scatter of $M/L_r$ in the corresponding regions. 
We tested in previous work \citep{Zhu2018a} that assuming a constant mass-to-light ratio (even if it actually varies) to infer the gravitational potential does not affect the stellar orbit distribution in the dynamical model \citep{Zhu2018a, Poci2019}, and thus does not affect the luminosity fraction of hot inner stellar halo $f_{\rm halo}$.
Furthermore, the $\Mstar/L_r$ ratios from stellar population synthesis in the halo-dominated outer regions are roughly consistent with the $\Mstar/L_r$ ratios inferred from our dynamical model. 
With $\Mstar/L_r = 3.5\pm0.5$ \MLsun, we have $\Mshalo = (4.4\pm0.8)\times 10^{10}$\,\Msun, which is still within $1\,\sigma$ of our quoted reference value.

\section{A NGC 1427-like galaxy in TNG50}
\label{A:ngc1427-ana}

NGC 1427 does not have a significant cold disk and its hot inner stellar halo is younger than the bulge, which is different from NGC 1380. While the galaxies we chose for case studies in \citet{Zhu2021} are all NGC 1380 analogues, here we present a NGC 1427-like galaxy taken from TNG50. 
About $25\%$ of TNG50 galaxies have hot inner stellar halos younger than bulges. One of these is TNG50 465255 with $M_*=9.3\times10^{10}$\,\Msun\,, $R_e=4\,$kpc, and a hot inner stellar halo mass of $\Mshalo=1.2 \times 10^{10}$\,\Msun. The galaxy has experienced a massive merger started at $t\sim 6$ Gyr and ended at $t\sim 4.8$ Gyr with a stellar mass accreted from the secondary progenitor of $M_{*,\rm Ex1}= 1.7 \times 10^{10}$\,\Msun. No other merger more massive than $0.1 M_{*,\rm Ex1}$ happened after that. The total ex-situ stellar mass is $M_{*,\rm ExSitu}= 2.2 \times 10^{10}$\,\Msun. 

\begin{figure*}
\centering\includegraphics[width=16cm]{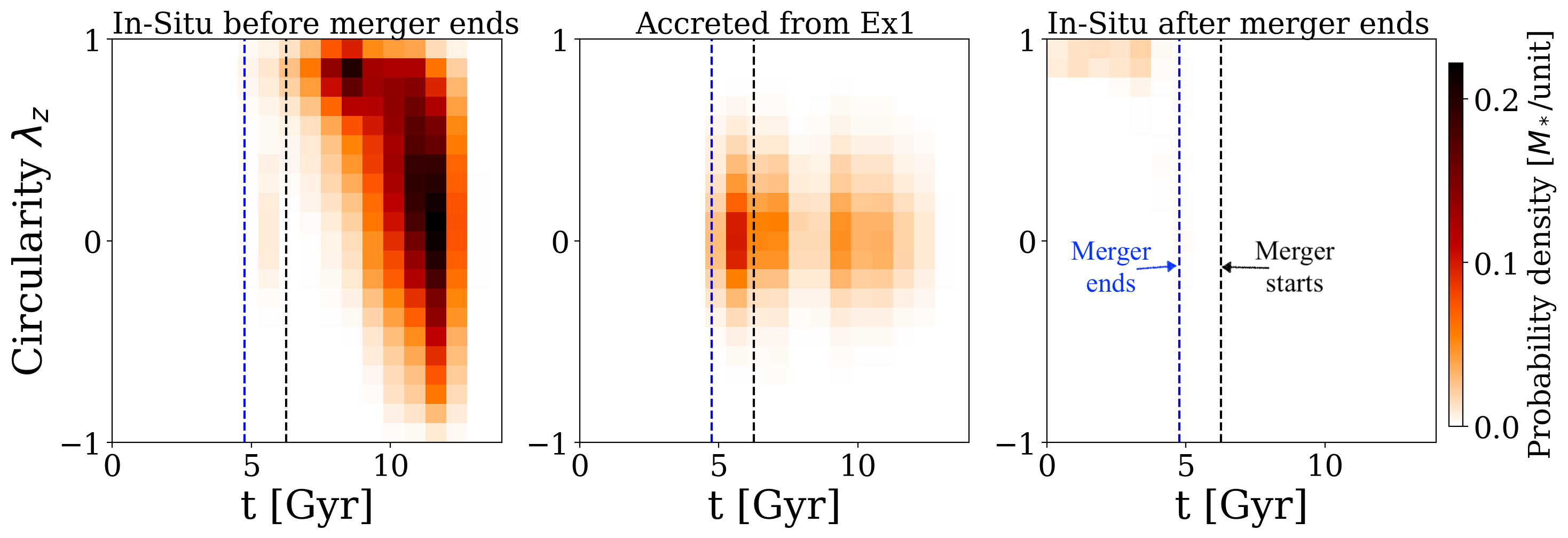}
\caption{
  \textbf{Probability density distributions in phase space of stellar age $t$ versus circularity $\lambda_z$ for stars with different origins in TNG50\,465255.} From left to right, they are in situ stars formed before merger ended, stars accreted from the secondary progenitor, and stars formed in situ after the merger ended. The black and blue dashed lines indicate the merger starting and ending time.     
 }
\label{fig:tlz_465255}
\end{figure*}

 We split the stars in the galaxy at $z=0$ into three categories according to their origins: (1) stars formed in situ before the merger ended, (2) stars accreted from the secondary progenitor, and (3) stars formed in situ after the merger ended. 
 In Figure~\ref{fig:tlz_465255}, we show the probability density distributions in phase space of stellar age $t$ versus circularity $\lambda_z$ for each part. 
The in situ stars formed before the merger ended have a wide range of $\lambda_z$ in the galaxy at $z=0$. The dynamical hot stars contribute mostly to the bulge and they are very old. The stars accreted from the secondary progenitor are younger and mostly contribute to the hot inner stellar halo. After the merger ended, only very few stars formed on dynamically cold orbits with little gas left after the merger. At the end, TNG50\,465255 is an elliptical galaxy.

\begin{figure*}
\centering\includegraphics[width=16cm]{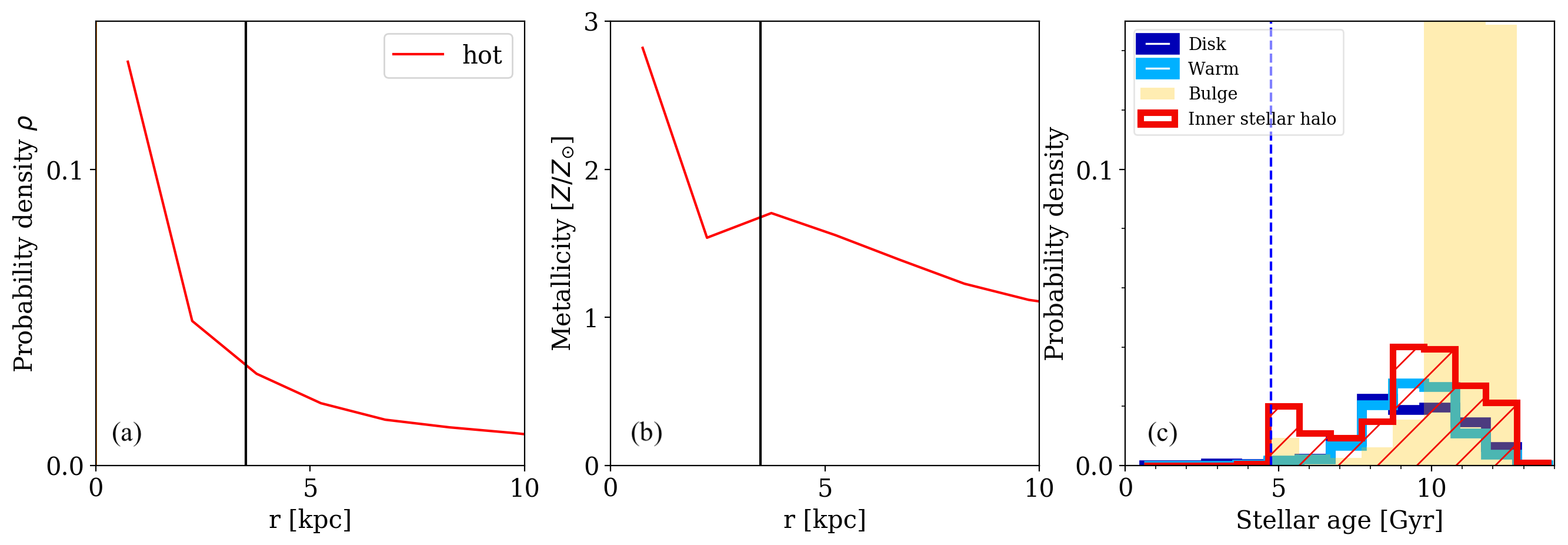}
\caption{\textbf{Some internal properties of TNG50\,465255}. Panel (a): probability density of stars on dynamical hot orbits with $\lambda_z<0.5$ as a function of radius, Panel (b): Metallicity distribution of stars on the hot orbits as a function of radius. Panel (c):
 Stellar age distribution of different components. Stars in the hot inner stellar halo are younger than those in the bulge, similar to what we found for NGC 1427.
 }
\label{fig:tdis_465255}
\end{figure*}

We separate the galaxy into four components: cold disk, warm component, bulge and hot inner stellar halo in the same way as we did for NGC 1427.
In the left panels of Figure~\ref{fig:tdis_465255}, we show the probability density and metallicity distribution of the hot orbits (with $\lambda_z<0.5$) as a function of radius. Here we use radial bin size of $\sim1$ kpc to avoid the dip of probability density at the inner-most bin. We separate the bulge and hot inner stellar halo at $r=3.5$\,kpc as we did for all the galaxies \citep{Zhu2021}, although the transition in density is not sharp and the metallicity has already flatten at smaller radius, similarly to NGC\,1427.
The stellar age distribution of the four components is shown in the right panel of Figure~\ref{fig:tdis_465255}. Similar to NGC 1427, TNG50\,465255 has a younger inner hot stellar halo than the bulge. For this galaxy, the age of the youngest star in the hot inner stellar halo indicates the merger ending time.

\end{document}